\documentclass[twocolumn]{aastex63}
\usepackage{amsmath}

\newcommand{\imc}{\texttt{imcascade}}
\defcitealias{Cappellari2002}{C02}

\received{XXX}
\revised{YYY}
\accepted{ZZZ}

\submitjournal{ApJ}

\shorttitle{Baysian MGE models with \imc}
\shortauthors{Miller \& van Dokkum}

\graphicspath{{./}{figures/}}

\begin{document}

\title{Bayesian fitting of multi-Gaussian expansion models to galaxy images}

\correspondingauthor{Tim B. Miller}
\email{tim.miller@yale.edu}

\author[0000-0001-8367-6265]{Tim B. Miller}
\affiliation{Department of Astronomy, Yale University, New Haven, CT 06511, USA \\}

\author[0000-0002-8282-9888]{Pieter van Dokkum}
\affiliation{Department of Astronomy, Yale University, New Haven, CT 06511, USA \\}

\begin{abstract}
Fitting parameterized models to images of galaxies has become the standard for measuring galaxy morphology. This forward modelling technique allows one to account for the PSF to effectively study semi-resolved galaxies. However, using a specific parameterization for a galaxy's surface brightness profile can bias measurements if it is not an accurate representation. Furthermore, it can be difficult to assess systematic errors in parameterized profiles. To overcome these issues we employ the Multi-Gaussian expansion (MGE) method of representing a galaxy's profile together with a Bayesian framework for fitting images. MGE flexibly represents a galaxy's profile using a series of Gaussians. We introduce a novel Bayesian inference approach which uses pre-rendered Gaussian components, which greatly speeds up computation time and makes it feasible to run the fitting code on large samples of galaxies. We demonstrate our method with a series of validation tests. By injecting galaxies, with properties similar to those observed at $z\sim1.5$, into deep HST observations we show that it can accurately recover total fluxes and effective radii of realistic galaxies. Additionally we use degraded images of local galaxies to show that our method can recover realistic galaxy surface brightness and color profiles. Our implementation is available in an open source python package~\imc, which contains all methods needed for the preparation of images, fitting and analysis of results.
\end{abstract}


\section{Introduction} \label{sec:intro}
The morphologies of galaxies hold clues about the physical process which shaped them~\citep{Conselice2014}. The size, often defined as the radius containing X\% of a galaxy's light or mass, and the surface brightness profile, often characterized by the Sersic index~\citep{Sersic1968} or ratio of radii, can help distinguish between different formation pathways and different models of galaxy formation ~\citep{Carollo2013,Whitaker2015,VanDokkum2015, Suess2021}. Galaxy morphology has been extensively studied focusing on various types of galaxies, using various instruments to help produce a consistent picture of galaxy formation from low to high redshift~\citep{Shen2003,Trujillo2006, Williams2010,Ono2013,vanderWel2014,Lange2015,Mowla2019}.

\subsection{Parameterized Models}

The modern standard for measuring galaxy morphology involves fitting parameterized models directly to observed images. This forward modelling approach allows one to account for the point spread function (PSF), sky background and other systematics while using empirically-motivated functions to describe the radially-varying intensity.  Some of the most commonly used fitting codes are \texttt{GALFIT}~\citep{Peng2010}, \texttt{IMFIT}~\citep{Erwin2015} and \texttt{ProFit}~\citep{Robotham2017}. The most commonly used function to describe the light distribution is the Sersic profile~\citep{Sersic1968}. Although all these codes allow the use of other functional forms as well as superpositions of different functions. There are also variants that explicitly focus on physically-motivated two component (bulge and disk) decompositions of galaxies such as \texttt{GIM2D}~\citep{Simard1998} and \texttt{BUDDA}~\citep{deSouza2004}. These codes, and others, have been successfully employed to study the morphologies of galaxies across a wide range of observational surveys.

While there are differences between the methods listed above, they all generally follow a similar outline. First a parameterization is chosen and the intrinsic model image is rendered. Then it is convolved with the PSF, often using Fourier transforms, to be compared to the observed images. This process is repeated many times as the fit parameters are optimized. A non-trivial aspect of these algorithms is choosing the parameterization and knowing if and when to add additional components. Many studies simply stick to a single component Sersic model \citep[e.g.][]{vanderWel2014,Lange2015,Mowla2019}. In galaxies that are composed of a bulge and a disk the Sersic index has been suggested as a proxy for the bulge-to-disk ratio. However, the index is not only sensitive to the relative luminosities of the bulge and disk but also to their relative sizes  ~\citep{Lang2014}. Furthermore, interesting features that only contribute a small fraction of the total luminosity, such as bars and stellar halos, are often lost as the parameterized model is more sensitive to higher S/N components. Another issue is that for galaxies with high Sersic indices ($n\gtrsim 4$) the model profile has large wings, such that a large fraction of the total light is at radii where the profile has a very low S/N ratio (or is not even included in the fit). More generally, assessing uncertainties in parameterized models is difficult due to unknown systematics, when comparing to the true light profile of galaxies.

\subsection{Non-parameteric Morphology}

In addition to fitting paramaterized models, there are ways to study the non-parametric morphology of galaxies. These include the CAS (Concentration - Asymmetry - Clumpiness) system~\citep{Abraham1996,Conselice2003} or direct measurements of clumps and other features from the resolved light or mass distributions ~\citep[e.g.][]{Zibetti2009,Wuyts2013}.  Similarly, one can measure the the average surface brightness profile using elliptical isophotes~\citep{ellipse, Stone2021}. Since these methods directly use observed images they are only applicable to galaxies which are much larger then the size of the PSF. In cases where the PSF is important one can attempt to perform deconvolution \citep[in one or two dimensions; see, e.g.,][]{vanDokkum2010,Shibuya2021}, but as deconvolution is not a unique process and tends to enhance noise the results can be difficult to interpret.

In an attempt to bridge parametric and non-parametric methods, \citet{Szomoru2010} introduced an extension of the Sersic fitting method. In their method a ``residual'' profile is created by subtracting the best fitting Sersic model, convolved with the PSF, from the observations. By simply adding this profile to the best fit unconvolved 1D Sersic profile, this method accounts for deviations from a Sersic model, especially on large scales. It works very well in a regime where the S/N ratio is high (such as HST images of moderate redshift $\sim L_*$ galaxies) and the galaxies are isolated (so that the radially-averaged residuals are not contaminated by other objects). \citet{Szomoru2012} This method has been used to confirm the compactness of $z\sim 2$ quiescent galaxies and to measure the color profiles of
galaxies \citep{Szomoru2012,Szomoru2013}. We refer to \citet{Suess2019} for an in-depth discussion of extracting color gradients from PSF-convolved HST
images using this and other methods.

\subsection{Approximating the Light Profile with a Series of Gaussians: The Multi-Gaussian Expansion method}

An alternative to fitting a parameterized model, or measuring the radial light profile directly, is to approximate a galaxy's light distribution with a series of Gaussians. Gaussians have special properties:  a series of Gaussians can reasonably approximate any other function; the convolution of two Gaussians is also a Gaussian; and the deprojection of a two-dimensional Gaussian is a three-dimensional Gaussian. These properties are very useful for astronomy, as convolution is needed to account for the effect of the PSF and deprojection is needed when studying the three-dimensional structure of galaxies. Due to the shape of a Gaussian, each component provides a localized contribution to the profile near its standard deviation. The concept of using a series Gaussians to represent galaxy's light profile was was first developed by \citet{Bendinelli1991} to deproject and deconvolve 1-D profiles. This represents a flexible view of a galaxies light distribution as it can represent any realistic profile without having to explicitly choose a physical model (such as Sersic or bulge+disk) a-priori .This was later expanded on by \citet{Monnet1992}, \citet{Emsellem1994a} and \citet{Emsellem1994b} who generalized the method to two dimensions and showed it could be used to accurately fit images of galaxies. These early studies focused on using a Gaussian representation to study a galaxy's deconvolved and deprojected mass distribution in tandem with kinematic information.

Building on these principles, \citet[][herafter,~\citetalias{Cappellari2002}]{Cappellari2002} developed a method to automatically fit Multi Gaussian Expansion (MGE) models to galaxy images\footnote{$\texttt{mgefit}$ is available here: https://pypi.org/project/mgefit/}. This method iteravely adds Gaussian until they no longer improves the fit. The method bins pixels along sectors in the outer parts of galaxies in order to improve the computational speed. This implementation is a powerful approach to studying galaxy surface brightness profiles. Especially when combined with Jeans modelling of kinematics~\citep{Cappellari2008}, it has been successfully employed to study wide range of topics, including the stellar initial mass function~\citep{Cappellari2012}, the fundamental plane of galaxies~\citep{Cappellari2013}.

Many other studies have used a series of Gaussians to approximate other parameterizations, specifically exploiting some of their useful properties. \citet{Hogg2013} suggest using a mixture of Gaussians to approximate exponential and de Vaucouleurs profiles to provide quick and more accurate convolutions when fitting the models to images. This has since been expanded upon and implemented in the \texttt{tractor} software package~\citep{Lang2020}. \citet{Bundy2012} suggest a similar method to measure colors of galaxies across observations with vastly different spatial resolutions, and \citet{Sheldon2014} use Gaussians to represent parameterized galaxy profiles, following \citet{Hogg2013}, when performing shear measurements for weak lensing. \citet{Shajib2019} use Gaussians to represent observed light and mass profiles and provide a formalism to combine strong lensing and kinematics studies. These authors provide an analytic integral transform to efficiently transform any analytical profile into Gaussian components.

\subsection{This Paper: Bayesian fitting of MGE models}

In this paper we introduce a Bayesian method to fit MGE models to images of galaxies. Given the flexibility of MGE models, combined with Bayesian techniques, this allows for a full exploration of the uncertainties on morphological parameters free of systematic effects incurred by traditional parameterizations. In addition to these techniques, the decoupled nature of the components allows one to characterize the low surface-brightness outskirts of galaxies without systematic effects due to the high S/N central regions. This paper is organized as follows. In Section~\ref{sec:concept} we discuss the MGE concept and its usefulness when study galaxy morphologies. In Section~\ref{sec:implementation} we discuss our specific python implementation, \imc, which we make publicly available via github\footnote{https://github.com/tbmiller-astro/imcascade}, including our novel bayesian inference technique using pre-rendered images. Illustrative examples and validation tests are presented in Section~\ref{sec:tests}. Finally we summarize and discuss our method in Section~\ref{sec:summary}.

\section{Concept} \label{sec:concept}
\begin{figure*}
    \centering
    \includegraphics[width = 0.9\textwidth]{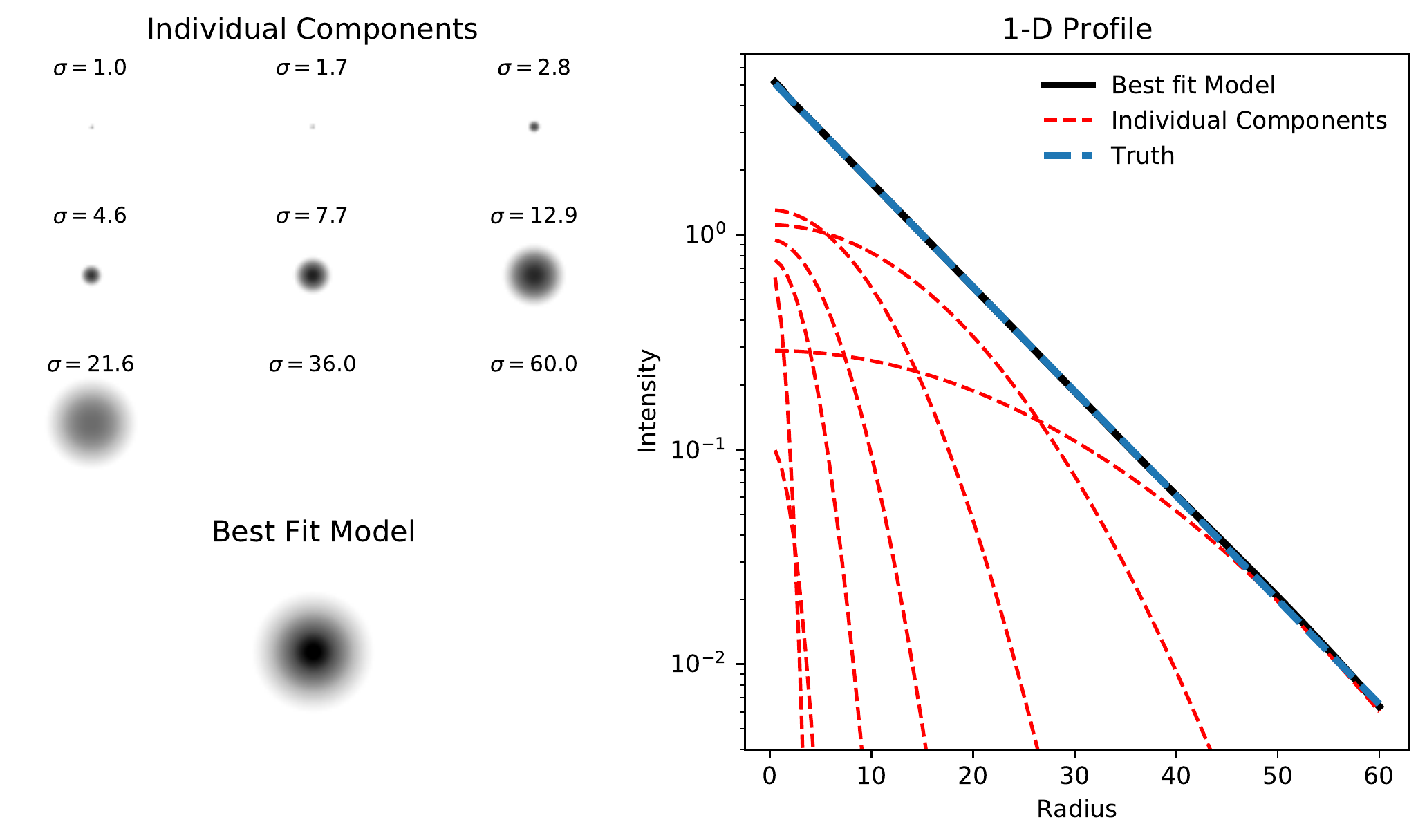}
    \caption{An example\ \imc~fit to an ideal exponential light profile is shown. Top left: The contribution of each of the nine individual components to the final best fit model in two dimensions. Bottom left: The final best fit model, which is the sum of the individual components. Right: The surface brightness profiles of the components and best fit model compared to the truth. In this particular fit the 5th, 6th and 7th components with widths of 7.7, 12.9 and 21.6 pixels respectively have the highest flux and dominate the light profile.}
    \label{fig:concept}
\end{figure*}

\subsection{Modelling the light distribution}
\imc~is application of the MGE formalism developed by \citet{Emsellem1994a,Emsellem1994b} and \citetalias{Cappellari2002}. Specifically, both the galaxy and PSF profile are represented by a series of Gaussians. The $i$'th Gaussian in the series is identified by its weight or flux, $a_i$, and width $\sigma_{g,i}$. This is also referred to as a Gaussian mixture model, where the ratios of fluxes are the mixing fractions. Throughout this paper we will be using the width of a Gaussian component to mean it's standard deviation. We assume that all the Gaussians have the same central position, $(x_0, y_0)$, axis ratio, $q$, and position angle, $\phi$.  The intrinsic projected light distribution of a galaxy, $G$, as modelled with $N$ components, is written in our notation as

\begin{equation}
\label{eqn:G}
    G(x,y) = \sum_{i=0}^N \frac { a_i }{2\pi\, q\, { \sigma_{g,i} }^2 }\ \exp\left[ - \frac{{x^{\prime} }^2 + \left( \frac{y^\prime} {q}\right) ^2}{2{\sigma_{g,i}}^2}  \right],
\end{equation}
with
\begin{equation}
    \label{eqn:prime_coord}
    \begin{aligned}
        x^{\prime}& = (x-x_0) \cos{\phi} + (y-y_0) \sin{\phi} \\
        y^{\prime}& = -(x-x_0) \sin{\phi} + (y-y_0) \cos{\phi}.
    \end{aligned}
\end{equation}

For simplicity we assume in the following that the PSF is circular, as in \citetalias{Cappellari2002}. However, convolution is also possible for two non-circular Gaussians~\citep{Monnet1992}, which has been implemented in \imc~(with a slight modification to the original derivation provided in Appendix~\ref{sec:NC_psf} ) . The PSF is represented as another series of $M$ Gaussians with widths, $\sigma_{p,j}$ and weights, $b_j$. Therefore the light distribution of the PSF, $P$, is similar to Eqn.\ \ref{eqn:G} with $q = 1$ and $\phi= 0$ for the circular case:

\begin{equation}
    \label{eqn:P}
    P(x,y) = \sum_{j=0}^M \frac{b_j}{2\pi\, {\sigma_{p,j}}^2} \exp \left[ - \frac{x^2 + y^2}{2{\sigma_{p,j}}^2}  \right].
\end{equation}

The observed light distribution, $O$ is then the convolution of $G$ and $P$. Convolution is distributive and the convolution of two Gaussians is another Gaussian. It follows that $O$ is a series of $N\times M$ Gaussians containing each component in $G$ convolved with each component in $P$, which are themselves Gaussians:

\begin{equation}
    \label{eqn:O}
    \begin{aligned}
            O(x,y) &=\ G(x,y) \bigotimes P(x,y) \\
            &=\ \sum_{j=0}^M \sum_{i=0}^{N} \frac{a_i\, b_j}{2\pi\, q_{i,j}\, {\sigma_{i,j}}^2} \ \exp \left[ - \frac{{x^{\prime} }^2 + \left( \frac{y^\prime} {q_{i,j}}\right) ^2}{2{\sigma_{i,j} }^2}  \right],
    \end{aligned}
\end{equation}
with $\bigotimes$ denoting convolution and
\begin{equation}
    \label{eqn:O+}
    \begin{aligned}
        \sigma_{i,j} ^2&= {\sigma_{g,i} }^2 + {\sigma_{p,j} }^2\\
        q_{i,j} ^2 &= \frac{ (q\, \sigma_{g,i} )^2 + {\sigma_{p,j} }^2  } {\sigma_{i,j} ^2 }.
    \end{aligned}
\end{equation}

The observed light distribution, $O$, is what will be compared to data to find the best fit model. When given a model for the PSF, one can analytically calculate the observed light distribution above, forgoing the need for traditional convolution techniques. This is a large benefit to MGE technique as numerical convolution, often performed using Fourier transforms, is computationally expensive and can introduce numerical errors. Here we can directly render the convolved model when comparing to data.

\subsection{Morphological Quantities} \label{sec:morph}
An important application of MGE models is the study of morphological quantities of galaxies. Given it's flexible representation of a galaxy's surface-brightness profile, the derived parameters are less biased than parametric models which do not accurately portray a realistic galaxy. MGE models were first used to study the fundamental plane and other morphological relations in ~\citet{Cappellari2013}. It has since been used in many other contexts to study galaxy morphology from low to high redshift ~\citep{Shetty2015,Hongyu2018,Mendel2020,Trevor2020,Deugenio2021}

As in \citet{Cappellari2013}, the cruve-of-growth (COG) for an MGE model where all the components have the same axis ratio is calculated as,
\begin{equation}
    \label{eqn:COG}
    f(< r) = \sum_{i=0}^N a_i \left( 1 - \exp\left[ - \frac{r^2}{2{\sigma_{g,i}}^2}\right] \right).
\end{equation}
The total flux of the MGE model is then calculated below.
\begin{equation}
    \label{eqn:flux}
    F = f(< \infty) = \sum_{i=0}^N a_i.
\end{equation}

There is no analytical solution for $r_{50}$ (or the effective radius) or any other fractional radius, but these values can be solved for quickly by interpolating the COG from a grid of radii values~\citep{Cappellari2013}\footnote{These equations are also implemented in the package jampy ( https://pypi.org/project/jampy/)}. All these basic calculations have been implemented in our package~\imc, along with other quantities such as the Petrosian radius \citep{Petrosian1976,Blanton2001}, the isophotal radius \citep{Redman1936,Holmberg1958,Trujillo2020} among others. While the analysis of morphological quantities in MGE models is slightly more complicated than parameterized models, which often directly fit for quantities like $r_e$ and the total flux, any quantity of interest can be calculated.

\section{Implementation}
\label{sec:implementation}

We have implemented \imc\ in a open source python software package which we make available here\footnote{https://github.com/tbmiller-astro/imcascade}. This package allows one to fit any astronomical source in two dimensions using the formalism discussed in Section~\ref{sec:concept}. Since using \imc\ requires a Gaussian decomposition of the PSF, which is not always available, we include a module to convert a pixelized PSF to a Gaussian decomposition. By default we use a tilted-plane to model the sky background simultaneously with the galaxy. We have implemented options for both $\chi^2$ minimization to find the best fit models and posterior estimation to estimate uncertainties. Additionally we have included a module to help analyze the results and calculate morphological quantities of interest (see Sec.~\ref{sec:morph}). In this section we describe the details of our implementation. For full documentation and source code please see our github repo.

\subsection{Model rendering}
\label{sec:render}

Standard parameterized fitting routines generally render the intrinsic galaxy image and then convolve this with a pixelized version of the PSF. This is typically done using Fourier transforms which are computationally expensive. Additionally there are limitations to the accuracy due to pixelization, both for rendering the PSF and for rendering the analytical profile. One of the major benefits of MGE models implemented in~\imc is that no computational convolution is necessary as it is done analytically following Eqn.~\ref{eqn:O} and Eqn.~\ref{eqn:O+}. Therefore we render the observed model, $O$, which can be calculated analytically given the input Gaussian widths and the model for the PSF, and have no need for any traditional convolution.

The standard approach to rendering model images is to evaluate the analytic profile at the central position of each pixel.  However issues arise for objects where the profile changes quickly on the order of a pixel size, such that the central position of the pixel is not an accurate assessment of the total flux of the pixel. In this scenario, oversampling the pixel grid \citep{Robotham2017} or an alternative approach where models are generated in Fourier space \citep{Hogg2013,Lang2020} can lead to more accurate model generation. Since \imc\ uses only Gaussians, we opt for an integration technique where we analytically integrate the light distribution over the area of a pixel to ensure accurate model generation. The computational cost is roughly equivalent to oversampling by a factor of 2. However, it is accurate for any Gaussian width, even those much smaller than a pixel, and for any central galaxy position, i.e., if the galaxy is not centered on a pixel.

The integral of a Gaussian on a cartesian grid (such as a CCD) is derived in Appendix B of \citet{Cappellari2002a}. A caveat is that integral can only by evaluated analytically for an un-rotated elliptical Gaussian. We address this by rendering all Gaussians  with $\phi = 0$ and then rotating them numerically, using the spline interpolation implementation in \texttt{scipy.ndimage.rotate}~\citep{scipy}. While this can introduce numerical errors, we have tested a variety of rotations for realistic galaxies and the total flux is always recovered to $< 1\%$. Additionally, to increase speed, we only use the integrated pixel approach for components with width $\leq 5$ pixels. At widths larger then 5 pixels there is essentially no difference between traditional rendering, i.e., evaluating at the center of the pixel, and our pixel integration technique, with differences in the fluxes of individual pixels smaller then 1 part in $10^4$. The results is a hybrid approach, where components that have small widths are rendered with the analytic pixel integrated technique for accuracy and the larger components are rendered traditionally for speed.

Once the model for the galaxy is rendered, there is an additional option to use a model for the sky background. This is an important step as inaccurate sky modelling can bias the results, especially in the low-surface brightness outskirts of galaxies (Fischer et al. 2017). The default of \imc~is to use a tilted plane which involves three parameters, the overall background level and the slopes in each direction, however this is optional. This allows the freedom to capture any systematic behaviour in the background of the image. If large cutouts ($>10\, r_e$) are used the sky should be well-separated from galaxy features. In our testing the slope is generally measured to be small, less than 1 part in $10^5$ per pixel. We have also implemented the option to use a constant background model. The alternative is to forgo modelling the sky simultaneously by measuring and subtracting it before fitting. However this method can cause systematic issues as the outskirts of galaxies or nearby bright stars can bias the sky measurement affecting the results of the profile fitting (Fischer et al. 2017). Furthermore, the uncertainty in the sky value is then not taken into account in the profile. We recommend using the tilted-plane sky model in most circumstances when fitting galaxies with \imc, however it is optional.

\subsection{Fitting a model to data}
\label{sec:ls}

The following free parameters are fit. First, for every model there are 4 structural parameters: the central position of the galaxy, $(x_0,y_0)$, the axis ratio, $q$, and the position angle, $\phi$. These are the same for every Gaussian component. In addition to these, the flux of each Gaussian component, $a_i$, is fit. Our decision to fix the axis ratio and position angle is based on our intended use case of faint and semi-resolved galaxies for which variations in these parameters with radius are often not discernible. However, we plan to add the option to allow these to vary in the future to better model well-resolved, high S/N galaxies. The number of Gaussian components and their widths is chosen beforehand, for more details see Section~\ref{sec:widths} below.  Finally if a tilted-plane sky model is used there are three additional free parameters. The total number of free parameters to be optimized is then $N_{\rm free} = 4 + N_{\rm Gauss} + N_{\rm sky}$.

We implement a $\chi^2$ minimization routine to optimize the free parameters and find the best fit solution. In particular we employ the Trust-Region Reflective \citep{trf} algorithm implemented in \texttt{scipy.optimize.least\_squares} routine~\citep{scipy}. This is inherently a bound-constrained method which effectively explores the entire parameter space. This is useful for our implementation since the fluxes of each component, $a_i$, do not map simply to the surface brightness profile, radii, or any other quantity that can be estimated a-priori. Therefore it is difficult to reliably guess the initial values based on observed properties of the galaxy, and other algorithms would risk getting stuck in local minima. The code is designed to accept a guess for initial values; however in practice we find the best fit parameters are relatively insensitive to changes in the initial guesses. To find the best-fit parameters $\chi^2$ is minimized:
\begin{equation} \label{eqn:chi2}
    \chi^2 = \sum_{i:\rm pixels} w_i\, \left( D(x_i,y_i) - O(x_i,y_i) \right)^2,
\end{equation}
with $D(x_i,y_i)$ the observed image, $O(x_i,y_i)$ the model profile described by Eqn.\ \ref{eqn:O}, $w_i$ the pixel weight. The weight is typically given as $w_i = z_i / {\sigma_{i} }^2$, where $z_i$ is equal to $0$ or $1$ and describes the pixel mask and ${\sigma_{i} }^2$ is the pixel variance.

In our implementation, the default is to explore the weights in logarithmic space i.e. exploring $\log a_i$ instead of $a_i$. We find that this is more accurate for the weights of the components with the largest widths. These components represent the, often faint, outskirts of galaxies which are generally below the noise limit of astronomical images and therefore often unconstrained. In logarithmic space, since it implies a reciprocal prior distribution, the fluxes of these unconstrained components tend to be smaller which is consistent with the physical picture of galaxies. However, exploring the weights in logarithmic space also constrains the weights to always be positive. This is fine in most cases but there may be some galaxies for which it is not accurate, e.g. galaxies with a large core. When fitting these types of galaxies with \imc~it is better to explore weights in linear space to allow for negative weights.

\subsection{Posterior Estimation}
\label{sec:bayes}

Along with finding the best fit model, we are interested in understanding the uncertainties in the derived morphological quantities. Parameterized models can bias not only the measurements but also the uncertainties of morphological quantities if the chosen model does not accurately reflect the true light profile of the galaxy. MGE models employed by \imc are much more flexible and therefore are less susceptible to systematic errors due to the choice of model. Therefore the posteriors recovered with \imc~will better represent the random uncertainty due to the data quality rather than systemic errors.

We employ the Dynamic nested sampling method implemented in \texttt{dynesty}\footnote{https://github.com/joshspeagle/dynesty} \citep{Skilling2006,Higson2019,dynesty} to sample the posterior distribution and obtain uncertainties on derived morphological quantities. By default the Likelihood is calculated simply as $\ln \mathcal{L} \propto - 0.5\, \chi^2$, where $\chi ^2$ is given by Eqn.~\ref{eqn:chi2}. This approach is useful for out purpose as we are also interested in the Bayesian evidence, however it may not be optimal. \citet{MMbook} offer a critique of nested sampling techniques for Gaussian mixture models suggesting that they are inefficient due to the multi-modal nature of the posterior and large number of parameters.

The simplest implementation of Bayesian inference would be to let all the parameters vary, including the four structural parameters. The downside is that the computation time is long as it involves rendering model images following the procedure described above for every iteration. Instead we introduce a different, faster approach. We utilize the fact that the four structural parameters ($x_0$, $y_0$, $q$, and $\phi$) are generally optimized accurately and precisely by the least-squares minimization algorithm. The values  of the fluxes of each components, $a_i$, and the parameters representing the sky background are much more uncertain. Additionally, since these are the parameters that directly control the surface brightness profiles and the effective radius these are the most important to explore the posterior distribution of.

We introduce a slightly altered approach where the structural parameters are fixed to the values that were determined in the $\chi^2$ minimization while the weights and sky values are varied. The key advantage is that this allows us to pre-render the observed light distribution of each Gaussian component based on the best fit structural parameters (essentially performing the sum over $j$ but not $i$ in Eqn.~\ref{eqn:O}, with fixed $x_0$, $y_0$, $q$, and $\phi$). This results in a matrix representing the observed light distribution corresponding to each Gaussian component with each weight equal to unity. When multiplied by a vector containing a list of weights for each component and then summed together, one can calculate the observed light distribution $O(x,y)$ for the given set of weights. Since this calculation is a straightforward matrix multiplication it is much faster to execute compared to traditional rendering methods.

With the pre-rendered components the speed-up over traditional methods is roughly a factor of 100. With this speed increase performing Bayesian inference with \imc\ on a large set, $\gtrsim \mathcal{O} (10^2)$, of galaxies becomes tractable. The obvious downside is the assumption that  the structural parameters found by least-squares minimization are accurate and that their uncertainties do not contribute significantly to the uncertainty in derived parameters such as the effective radius. We have tested these assumptions using injection and recovery tests in Section~\ref{sec:ir}.

A reasonable, albeit relatively simple, choice of priors for the component weights are log-uniform priors. As discussed above, since this implies a $1/a_i$ prior distribution, this provides more accurate results for the largest width components, which are often poorly constrained by the data. The reciprocal prior distribution constrains theses components to have smaller flux, which matches our physical picture of galaxies, which have faint outskirts. However, these cause issues in practice. Specifically when using nested sampling, the size of the prior directly influences the performance of the sampling. In our testing, the log-uniform prior were often too broad and hindered performance both in terms of speed and reproducability.

An alternative, and the default we use, is to use informed priors based on the results of the least-squares minimization. Specifically, we use a Gaussian prior for each component centered on the best-fit value with variance equal to 4 times the variance estimated during the least squares minimization routine. This variance is estimated using the derived Hessian matrix. In our testing, we find these informed priors yielded better performance, in terms of run time and reproducability, when using nested sampling for Posterior estimation. This approach is also referred to as Empirical Bayes and can lead to falsely overconfident results~\citep{Gelman2013}. To ensure our prior choice does not affect the Posterior we have run a series of tests on simulated galaxies in Sec.~\ref{sec:tests}. Additionally we have tested increasing the factor to expand the Hessian variance and found little change in the posterior distribution.

\subsection{Choosing the Set of Gaussian Widths}
\label{sec:widths}

One of the important choices that need to be made when running \imc~ is the number and widths of the Gaussian components. Similar to \citetalias{Cappellari2002}, the standard choice is to use 8 -- 10 components with widths logarithmically spaced from roughly 1 pixel (or half the PSF FWHM for critically-sampled data) to 10 times the estimated effective radius. Logarithmically spaced components allocates more components (and thus more flexibility) in the center of the galaxy, where the signal is the strongest and the light profile changes the fastest. The outer parts of the galaxy, where the signal is lower and the profile changes is shallower, require less components to model accurately. \citetalias{Cappellari2002} show that for a power law profile logarithmically spaced widths are the optimal choice. Since many galaxy profiles can be approximated by a broken power law, this is a logical starting place, however, it is not necessarily the optimal spacing in all cases.

While these choices matter, small changes within reason should not greatly affect the fit. For example, we have had success accurately modelling galaxies with widths equal spaced in hyperbolic arcsine space, which is similar to logarithmic scaling. Additionally, varying the number of components between eight and ten should not make a large difference. Systematic issue can arise if the widths are not chosen well. For example, if the width of the largest component is not large enough the total flux will likely be underestimated, as the light in the outskirts cannot properly be accounted for. Also, as we show below in Sec.~\ref{sec:tests}, choosing too few components can lead to inaccuracies in the recovered profile. We have implemented a function which uses the observed Kron radius to estimate the half-light radius and choose a set of components given only the galaxy and PSF images.~\citep{Kron1980}

Since the choice of widths when fitting is not unique it is possible to compute the joint probability distributions of derived morphological quantities from separate models which use a different set of widths. This is particularly useful when analyzing surface brightness profiles, as the discreteness of the series of Gaussians can cause artificial ``wiggles'' in the profile \citepalias[see][and Fig.~\ref{fig:exp_num_comp} below]{Cappellari2002}. Combining models with different widths can help smooth these artificial wiggles. To combine different models (i.e.fits to the same galaxy with different number of, or widths of, the components) we employ a Bayesian Model Averaging approach\citep{Roberts1965,Fragoso2018}. For each model the posterior and Bayesian evidence are calculated independently using the method described above. Then the joint posterior of all the models for any morphological quantity, like flux or $r_{50}$ for example, is the weighted sum of the posterior for each model. The weight for each model is the ratio of the Bayesian evidence for that model to the model with the maximum Bayesian evidence. Since our choice of widths for the Gaussian components is not unique, we simply use an equal prior for each model. Combining models in this way to produce joint posteriors lessens the impact of a specific choice of widths and can help with discretization errors caused by approximating the profile with a discrete number of Gaussians.

A complimentary approach to model averaging is the use of the Bayesian Information Criterion \citep[BIC][]{Kass1995} to decide how many components to use. This approach is common in Gaussian Mixture Models and is similar to \citetalias{Cappellari2002} who only add components if it decreases the relative error of the fit.

\section{Tests and Examples}
\label{sec:tests}
\subsection{Illustrative example}
As an initial illustrative example we fit a perfect exponential profile (Fig.\ \ref{fig:concept}). We highlight the individual components and how each contributes to the best-fit model. We use nine components with logarithmically spaces spaced widths from 1 pixel to 60 pixels. The exponential profile has a half-light radius of 15 pixels, and the fifth, sixth and seventh components with widths of 7.7 pixels, 12.9 pixels and 21.6 pixels respectively have the largest flux and therefore contribute the strongest to the fit. When using a fixed set of widths for the Gaussian components we expect that not every component will contribute to every profile but the important point is that some combination of these components can represent any galaxy profile.

\subsection{Initial Tests and Validation}
\begin{figure}
    \centering
    \includegraphics[width = 0.95\columnwidth]{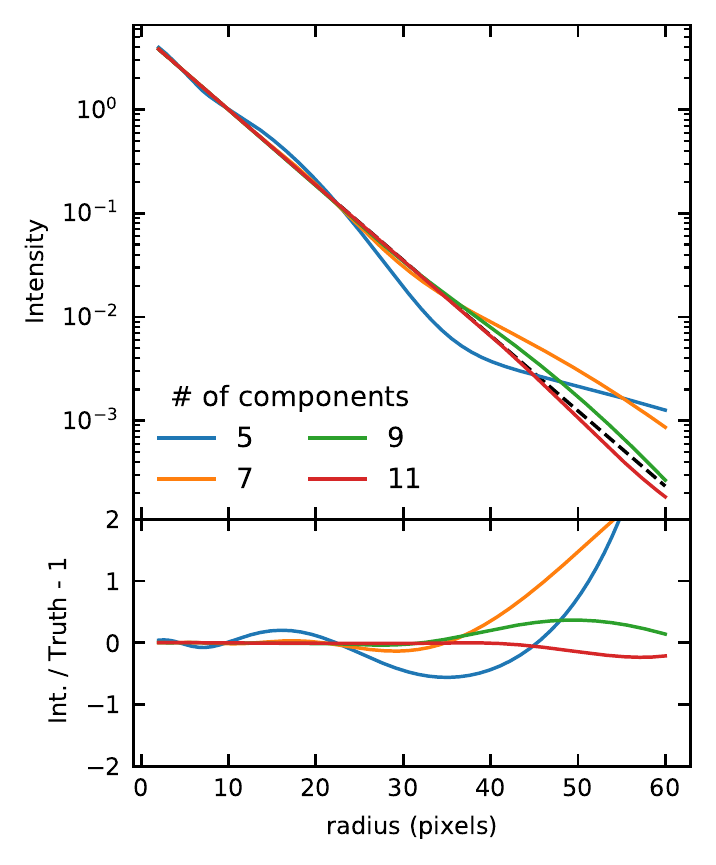}
    \caption{The best fit profiles for four \imc~models, with different number of components, which are fit to a perfect exponential profile are shown. The true exponential profile is shown as the black dotted line. Models with more components, recover the true profile with higher precision, with marginal improvement when increasing from nine to eleven components. This mirrors the precision of the recovered flux and effective radius}
    \label{fig:exp_num_comp}
\end{figure}

We run a series of tests and validation exercises using such simple exponential profiles. In the following  we use exponential profiles with $r_{\rm eff} = 10$ pixels centered in an image of $300 \times 300$ pixels and we fit them using \imc~with component widths logarithmically spaced from 1 pix to 95 pixels. This mirrors the set up for HST/ACS images of galaxies at $z\gtrsim 0.5$, with component widths logarithmically spaced from $\sim 1$\,pix to $10\times r_{\rm eff}$ with a cutout size of 30-40 times the effective radius.

We first show a precision test where we fit the perfect exponential profile without noise with 5, 7, 9, and 11 components. Surface brightness profiles based on the best-fit solution for each of the four models are shown in Figure~\ref{fig:exp_num_comp}. The \imc~model with five components shows large variations for the true exponential profile which increase at larger radii. There are pronounced wiggles in the profile at the locations of the component widths due to the small number of components. Even so, it is able to recover the total flux and effective radius to roughly 5\%, which holds true for various effective radii and profile shapes tested. Moving to seven components and nine components we find that the amplitudes of the wiggles dampen and the fluxes and effective radii are recovered more precisely to 1\% and 0.5\%, respectively. Again these results hold for all profiles tested. The results continue to marginally improve when considering 11 components, with roughly 0.1\% precision on the recovered flux and radii. However there are still some oscillations in the surface brightness profile which increase at larger radii. Given that the random uncertainties on the morphological measurements are almost always $>1$\%, the systematic errors when using more than seven components should be sub-dominant. Therefore, we suggest that using between 8 to 10 components when running \imc~for most cases as it provides precise results while not unnecessarily increasing the computation time.

\begin{figure*}
    \centering
    \includegraphics[width = 0.9\textwidth]{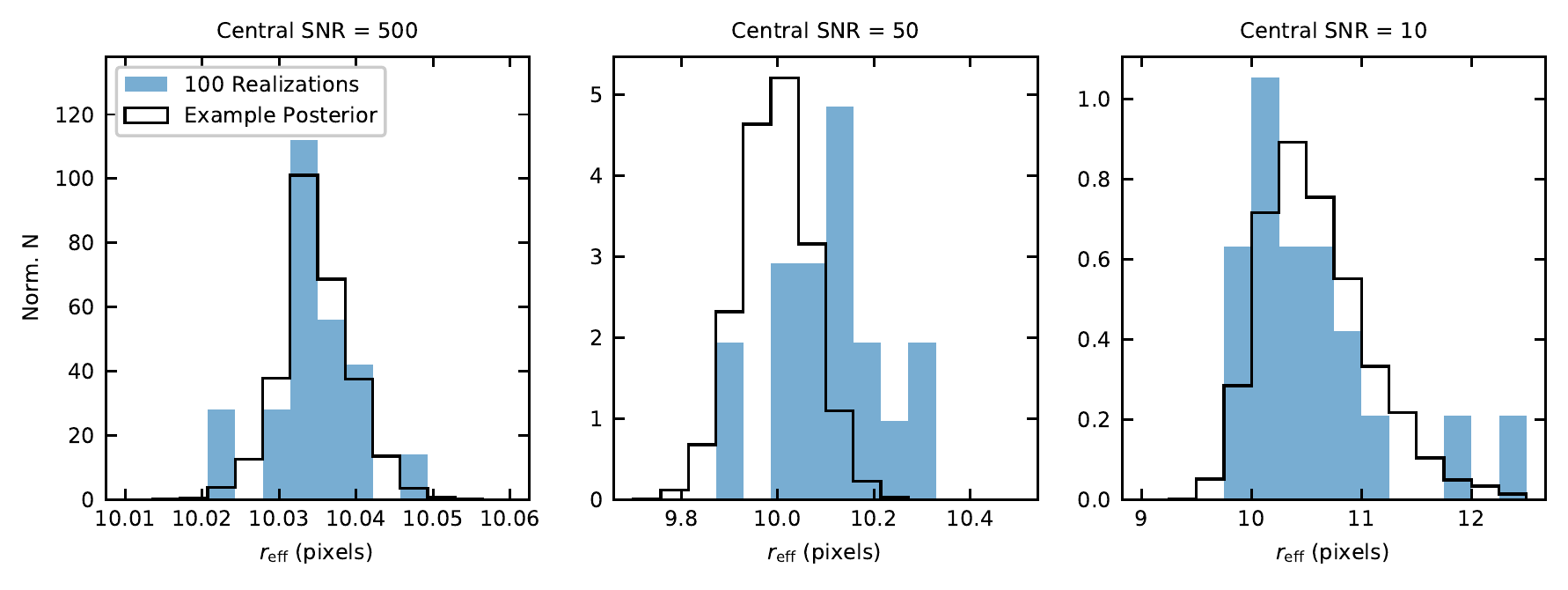}
    \caption{The effective radii of the best-fit \imc~model for an exponential profile with 100 noise realizations, for three noise levels. Also shown are the posterior distributions derived from our pre-rendered method from one of the example noise realizations. Over the three noise realizations, the shape, and width of the posterior distribution matches the distribution of best fit models for the 100 noise realizations.}
    \label{fig:exp_noise}
\end{figure*}

Next we introduce Gaussian noise to test how well our Bayesian techniques recover the uncertainties in the measurements. We introduce uniform Gaussian noise to each pixel to mimic sky noise; the tests are in the sky-limited (faint galaxy) regime and we do not include the Poisson noise for the model galaxies. To do this we perform least-squares fitting for a number of noise realizations and compare the distribution of radii from the best-fit models to the posterior distribution of one realization. Figure~\ref{fig:exp_noise} displays the results of these tests for three different levels of injected noise when using nine components. We find the distributions match well for each noise level. The widths and shapes of the example posterior and the set of 100 Monte Carlo realizations are consistent. This shows that the posterior distribution derived for our Bayesian approach represents realistic errors on the derived parameters. It is worth noting, that results from least squares minimization process tend to slightly overpredict the flux and effective radius. This is due to flux in the outer components being overestimated. These components are generally far below the noise and therefore difficult to constrain. Our Bayesian approach does a better job of capturing the uncertainties in these components and therefore produces more accurate results.

\subsection{Injection-recovery tests}
\label{sec:ir}
The next level of tests of \imc we perform are injection-recovery tests using real images. We inject fake galaxies into deep HST data to mimic galaxies at $z\sim 1.5$. We use the F160W data from the GOODS-S field from the CANDELS survey~\citep{Grogin2011,Koekemoer2011,Skelton2014}. We follow the observed properties of galaxies derived in \citet{vanderWel2012,vanderWel2014} and inject 100 fake galaxies using F160W magnitudes between 18\,--\,22 and effective radii between $0\farcs 3$\,--\,$0\farcs 9$. We model the galaxies as two-component Sersic profiles with a inner bulge component, with $3<n<6$, and an outer disk with $0.9<n<1.5$. The bulge radius is always smaller than that of the disk and we vary the fraction of light in the bulge, $f_b$, from 0 to 1. We convolve the fake galaxies using a pixelized version of the PSF. We run \imc\ on each galaxy using 10 Gaussian components, with widths logarithmically spaced between 0.5 pixels and $10\times r_{eff}$ using the pre-rendered method for posterior estimation. We model the PSF as a series of 4 Gaussians.

\begin{figure}
    \centering
    \includegraphics[width = 0.8\columnwidth]{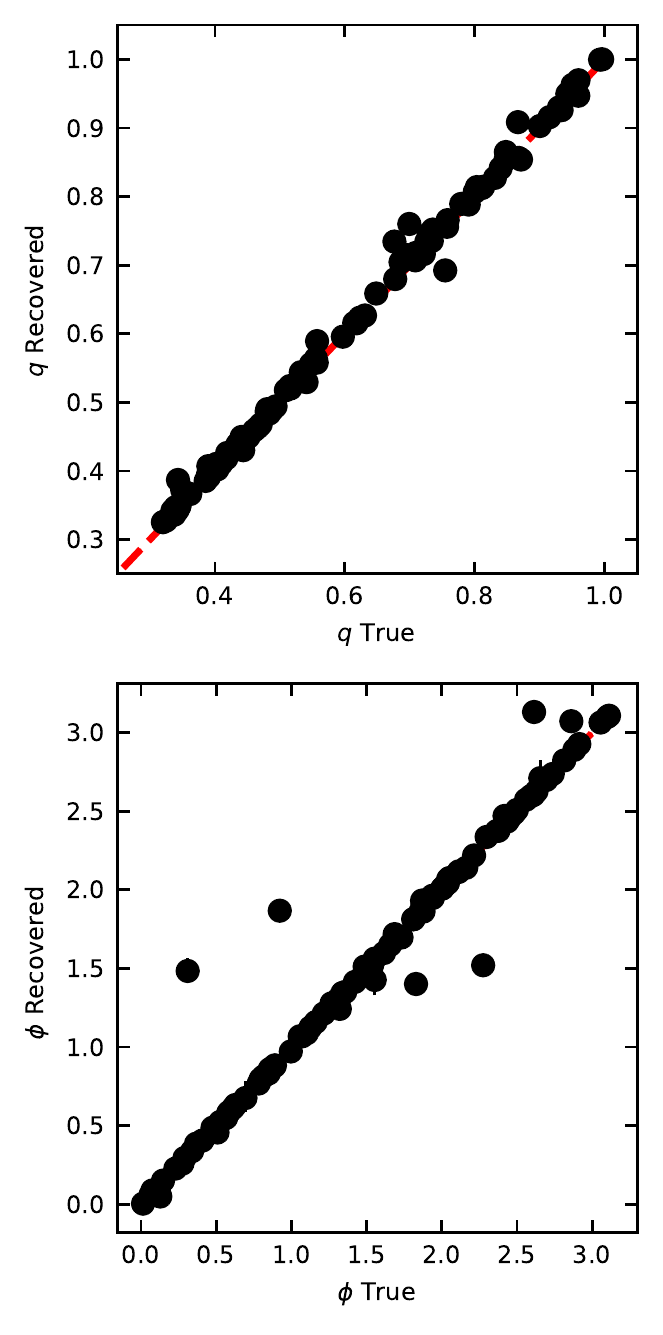}
    \caption{Displaying the best-fit axis ratio, $q$, and position angle, $\phi$ compared to the truth for \imc~models of $z\sim 1-2$ like galaxies injected into CANDELS data. The least squares routine consistently and accurately recovers the true axis ratio and position angle. The few apparent outliers in $\phi$ are very circular galaxies $q>0.95$ where the position angle is essentially meaningless and does not affect the fit. These results verify our use of these quantities when pre-rendering images to use in our ``express" Bayesian method.}
    \label{fig:IR_struct}
\end{figure}

In Figure~\ref{fig:IR_struct} we show the input axis ratio and position angle compared to those recovered from the least squares fitting. As explained above, these values are used to pre-render the images in the for our Bayesian inference technique. For the axis ratio, there is only a 2\% outlier fraction (2 of 100) with $| q_{\rm true} - q_{\rm recovered}| > 0.1$. There appear to be slightly more outliers when looking at the position angle; however most of these outliers are very circular, $q \gtrsim 0.95$, and therefore are unaffected by the offset in position angle. When accounting for these, there is a similar 2\% outlier fraction where the recovered position angle is offset by at least 10$^{\circ}$ from the truth. These results verify our assumptions that the least-squares minimization can accurately and reliably recover these structural parameters.

\begin{figure}
    \centering
    \includegraphics[width = 0.8\columnwidth]{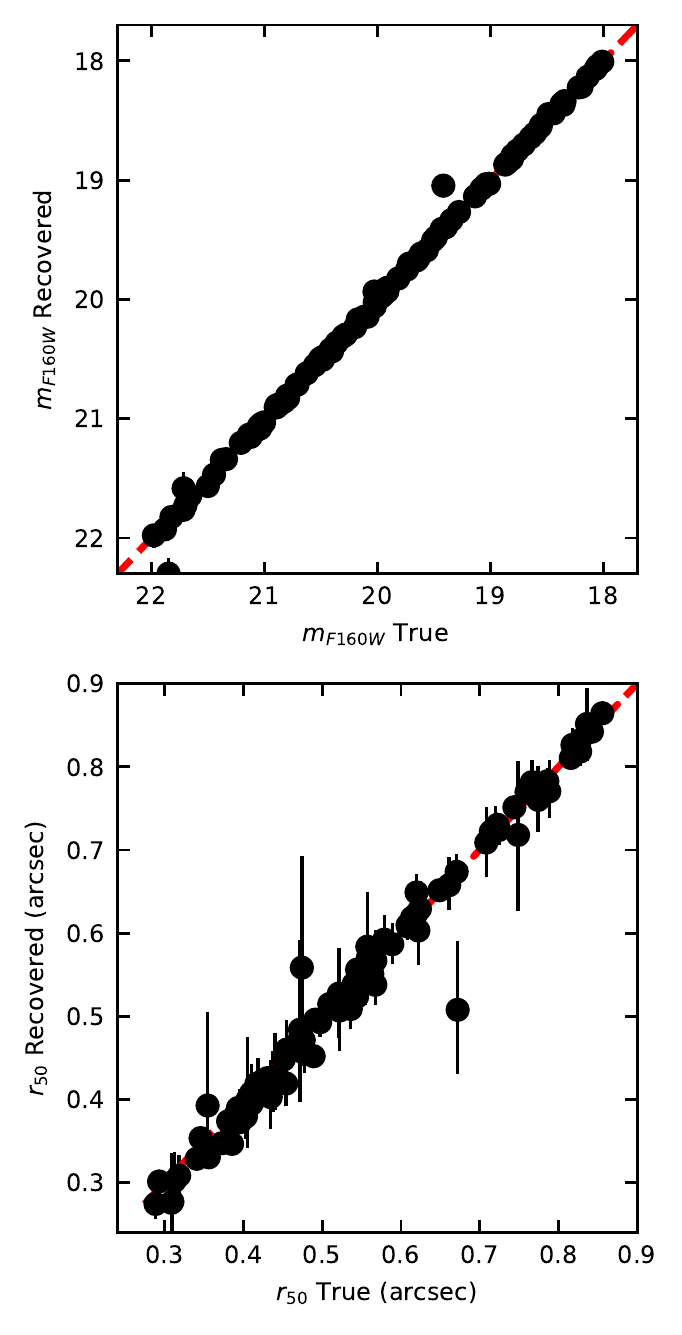}
    \caption{The total magnitude and effective radius measured using \imc\ are compared to the true values for our set of 100 injected galaxies. For each galaxy the errorbars represent the 16th-84th percentile range of the posterior. We find good agreement between the measured and true values displaying that \imc\ can accurately recover the morphology of $z\sim1.5$ like galaxies in CANDELS data.}
    \label{fig:IR_morph}
\end{figure}

The total magnitude and effective radii of the injected galaxies recovered using \imc~are shown compared to the truth in Fig.\ \ref{fig:IR_morph}. We find that \imc~is able to recover the flux and effective radii of these galaxies well in general. The total magnitudes are constrained extremely tightly with \imc. Of the 100 galaxies there is one galaxy with the recovered flux more then $5 \sigma$ away from the truth. Perhaps unsurprisingly this is correlated with and outlier in the structural parameters discussed above. Re-running galaxies using different input widths for the component and combining the models, as discussed in section~\ref{sec:bayes}, would likely help reduce this outliers. Similarly, the effective radii are well recovered with a similar outlier rate.

\subsection{Degraded images of local galaxies}

As an explanatory example of \imc\ in a realistic scenario we have used ground based data from the Hyper Suprime Cam Subaru Strategic Program (HSC-SSP) for local galaxies at $z \sim 0.01$ and degraded them so they resemble observations of galaxies at $z = 1-1.5$. This setup conveys how \imc~is able to recover the complexity of a true galaxy's profile.

Specifically we have queried the GAlaxy Mass and Assembley (GAMA) data release 3 database to find galaxies with $z = 0.01 - 0.015$ and stellar mass $\log M_* / M_\odot = 9.5 - 10.5$ and select 5 galaxies~\citep{Driver2011,Taylor2011}. We then obtained HSC imaging in the $g$ and $i$ band image from the HSC-SSP public data release 2 archive in a $2\arcmin \times 2\arcmin$ square around each galaxy\footnote{Downloaded using the \texttt{unagi} software package: https://github.com/dr-guangtou/unagi} ~\citep{Aihara2019}. We convolve each image with a Gaussian with $\sigma = 12$ pixels and then block average the images to degrade the sampling by a factor of 8. We then add additional uniform Gaussian noise equal to 0.05 ADU, roughly corresponding to 26 mag\,arcsec$^{-2}$ to mimic sky noise and decrease the S/N so that it is comparable to HST observations of galaxies at $z = 1-1.5$. This added noise dominates over that present in the original image. No additional Poisson noise is added. We first describe the fits to the degraded data, and discuss how well the Gaussian components are able to recover the surface brightness profiles of the degraded images. Then we compare the models to the original, undegraded profiles.

We fit each galaxy in the $g$ and $i$ band with 8 Gaussian components.  For each galaxy and image we run 5 models, in each the first Gaussian component has a width of 0.5 pixels and the remaining 7 are equally spaced in hyperbolic arcsine space with different start and end points. This is similar to logarithmic spacing altough slightly less steep, i.e. more components at larger radii. We use the pre-rendered method with \texttt{dynesty} to explore the Posterior with the Gaussian priors, based on the least-squares results explained in Sec.~\ref{sec:bayes}.

\begin{figure*}
    \centering
    \includegraphics[width = \textwidth] {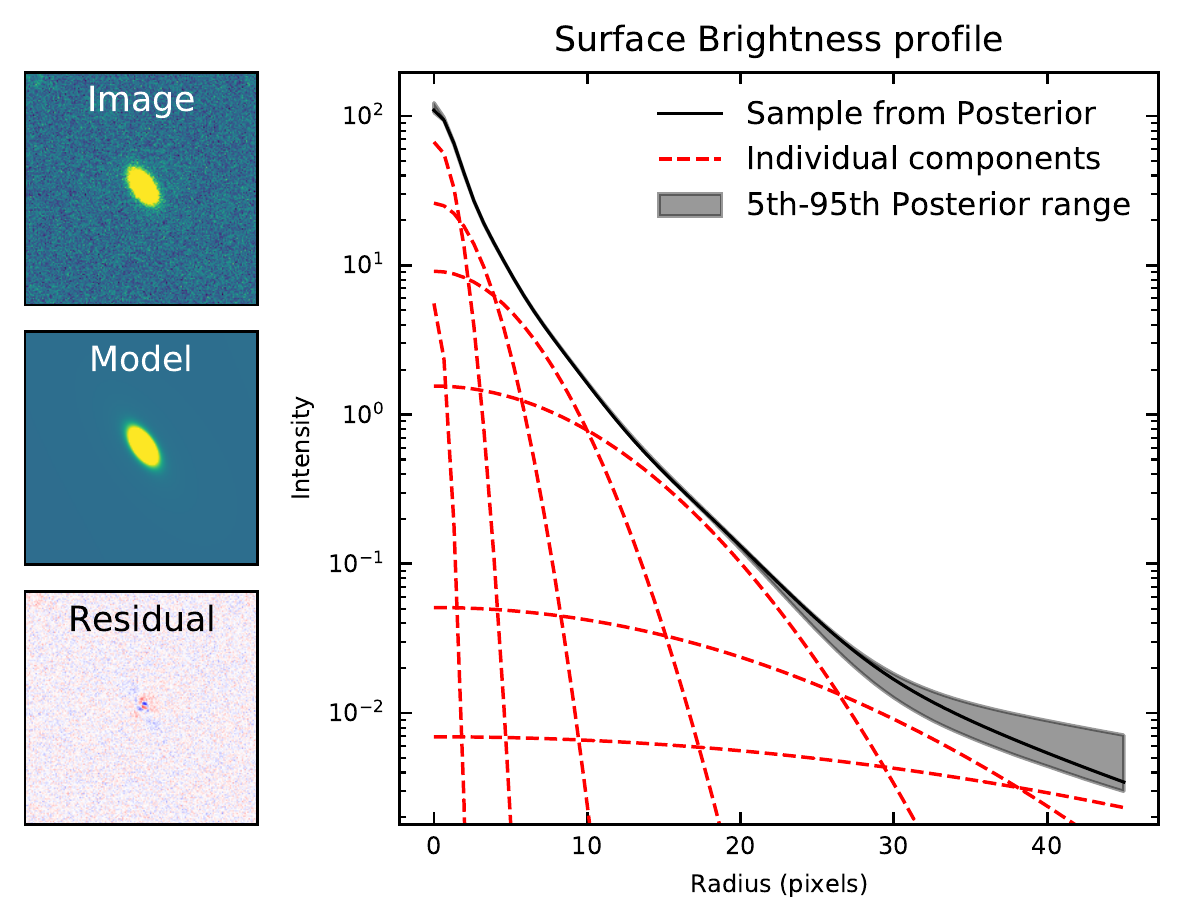}
    \caption{An example \imc~fit to a degraded image of a local galaxy is shown. Left: From top to bottom we show the original image, best fit model, residuals. Right: The intrinsic surface brightness profile recovered using our pre-rendered posterior estimation method is shown. We highlight the individual components of a single sample drawn from the posterier in red and show the 5th-95th percentile range of the total surface brightness profile in grey.}
    \label{fig:example_fit}
\end{figure*}

In Figure~\ref{fig:example_fit} we show an example of one of the degraded galaxies fit with \imc. We show the original image, the observed light distribution of the best fit model found using the least-squares minimization, and the residual image. We find that the best fit model matches the observed light distribution. The residual image has little structure and the residuals are Gaussian as expected. Also shown is the recovered surface-brightness profile. We show the 5th to 95th percentile range from the posterior distribution and highlight one sample from the posterior distribution, showing the individual components.

\begin{figure*}
    \centering
    \includegraphics[width = \textwidth]{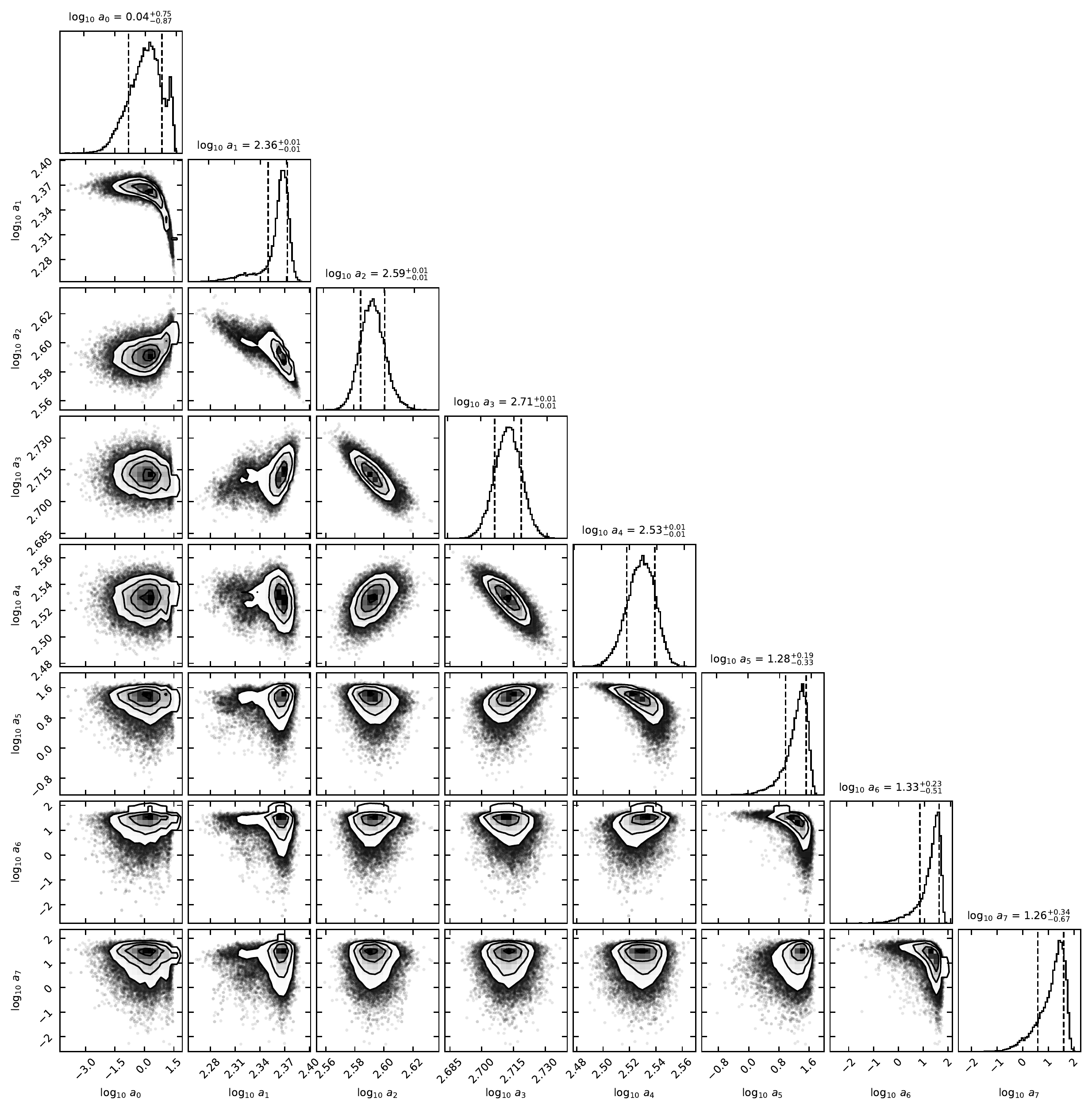}
    \caption{The joint posterior distributions for the weights of the 8 Gaussian components use to fit the Galaxy shown in Figure~\ref{fig:example_fit}. We find neighbouring components, i.e. those with similar widths, are often correlated as they contribute flux to the profile at similar radii. In this fit the 0th, along with 5th-7th components have much smaller weights and therefore do not strongly affect the profile.}
    \label{fig:post_weights}
\end{figure*}

A corner plot showing the posterior distributions of each of the Gaussian weights for the same galaxy is shown in Figure~\ref{fig:post_weights}. Again we see that some of the Gaussian components do not contribute strongly to this galaxy. In this case $a_0$ and $a_5$, $a_6$ and $a_7$ have much smaller weights compared to the rest of the components, while the 1st through 4th components have the largest weights. We also see that most of the weights have roughly Gaussian posterior distributions. However, the neighbouring components are often strongly anti-correlated with each other. These components have similar widths so they affect similar radii in the light distribution. If, for example the 3rd component is too low, then the 2nd and 4th components must compensate. It is also common to see components with the smallest widths be degenerate because when convolved with the PSF, they have similar widths.

\begin{figure}
    \centering
    \includegraphics[width = \columnwidth]{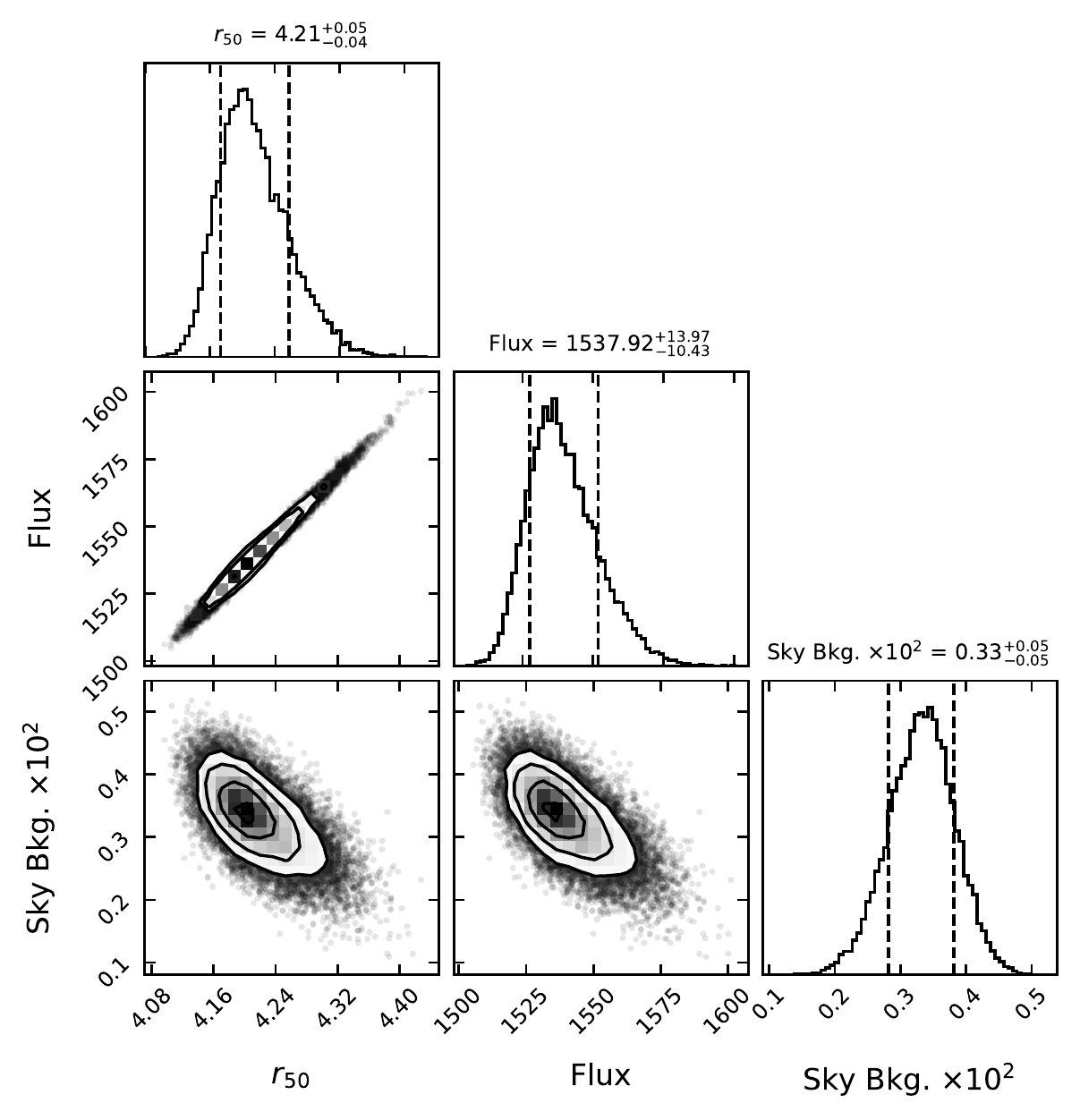}
    \caption{The posterior distributions of the flux and effective radius, $r_{50}$ and the  sky background level. The distributions of the flux and radii are  correlated, as expected. The uncertainty in both is largely driven by the uncertainty in the components with the largest width. Additionally both show covariance with the sky background level, highlighting it as an additional source of uncertainty on morphological parameters.}
    \label{fig:post_params}
\end{figure}

The posterior distributions of the derived quantities $r_{50}$ and flux are shown in Figure~\ref{fig:post_params}, along with the sky background. Both the radii and flux show roughly log-normal posterior distributions and they are highly correlated with each other. The uncertainties on these parameters are largely dependent on the uncertainties in the weights corresponding to the components with the largest widths, which are the least constrained. The models where these components have larger weights will correspond to larger flux and larger radii. The figure also displays that the sky-background is correlated with both  $r_{50}$ and the total flux, highlighting it as an additional source of uncertainty on morphological parameters.

\begin{figure}
    \centering
    \includegraphics[width = \columnwidth]{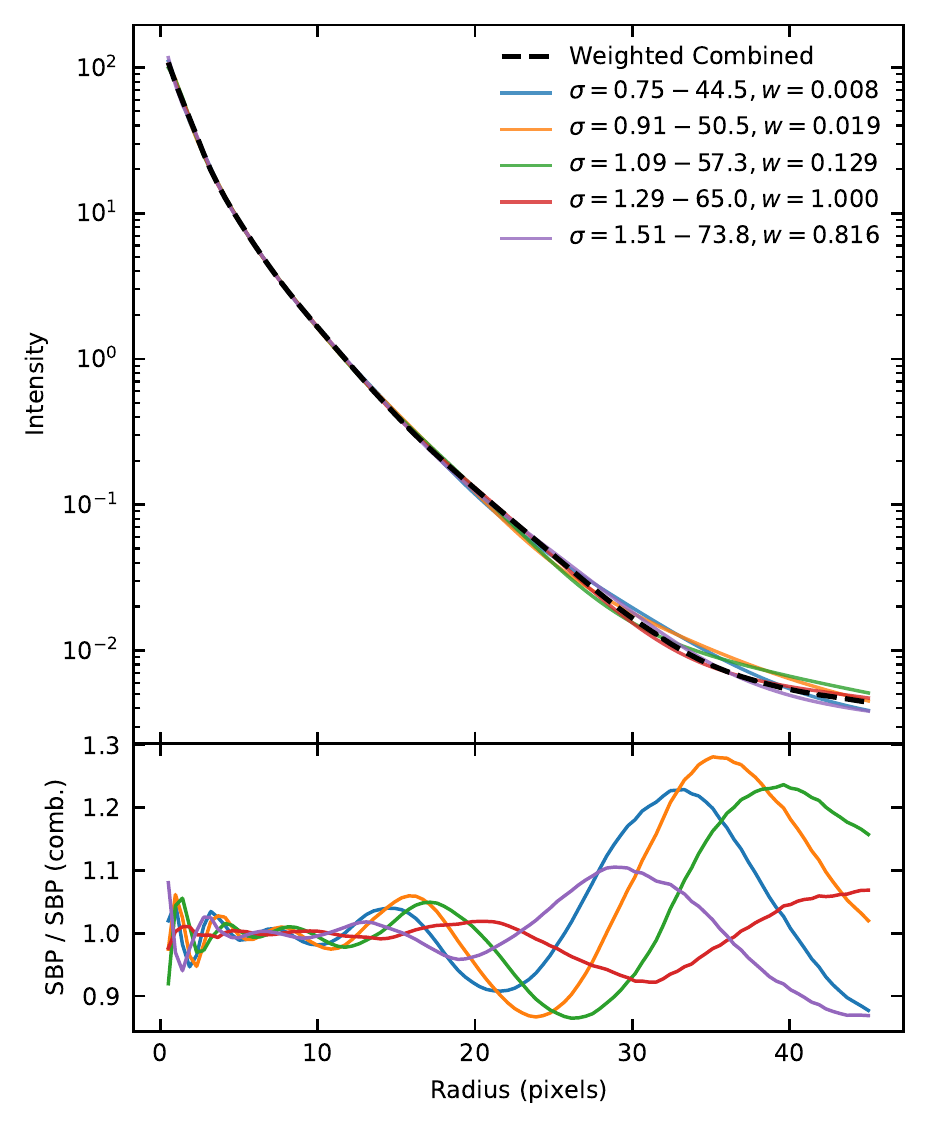}
    \caption{The median surface brightness profiles of five models, with different widths for the Gaussian components, are shown along with the combination weighted by the ratio of Bayesian evidences. The legend shows the relative weight of each model, with the highest weighted model shown in Fig. \ref{fig:example_fit},Fig. \ref{fig:post_weights} and Fig. \ref{fig:post_params}. The combined profile shows less "wiggles" due to discretization and our choice of the number of components.}
    \label{fig:comb_sbp}
\end{figure}

While in Fig.~\ref{fig:example_fit},~\ref{fig:post_weights} and ~\ref{fig:post_params} we have focused on using a single model using a single set of widths, as discussed in section~\ref{sec:bayes} it is possibly to independently explore the posterior distribution of several models each using different preset widths and combine the results. The surface brightness profiles for several models are shown in Figure~\ref{fig:comb_sbp}. Each set of widths has the same hyperbolic arcsine spacing but different start and end points, along with an additional component with a width of 0.5 pixels. The relative weight, $w$, of each model is shown and is calculated using the ratio of the Bayesian evidence to the maximum Bayesian evidence. We note that for this galaxy all of the models contribute somewhat, i.e. their relative weights are all non-zero, whereas for other galaxies one model dominates over the others.

\begin{figure}
    \centering
    \includegraphics[width = \columnwidth]{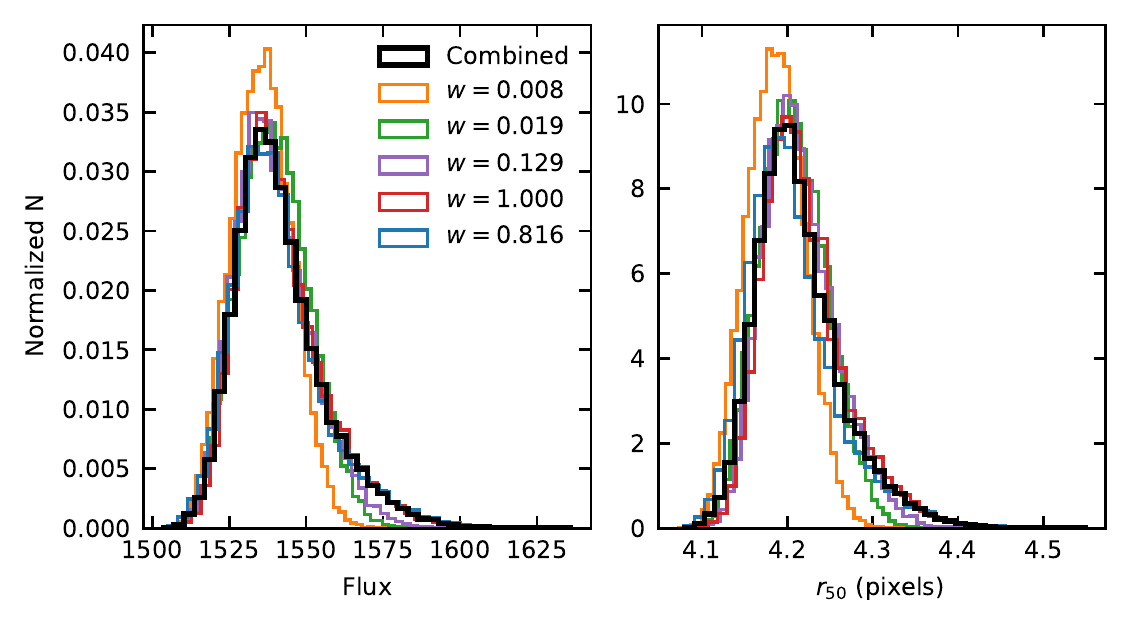}
    \caption{The posterior distribution for the flux and effective radii for the same five models shown in Figure~\ref{fig:comb_sbp} are shown along with the combined posterior distribution. All the median values of flux and $r_{50}$ for the five individual models agree with each other and the median combined value with 1-$\sigma$ demonstrating the insensitivity to these measurements on the choice of the widths for each component.}
    \label{fig:comb_params}
\end{figure}

When comparing the combined surface brightness profile to the individual models we find it is smoother with less ``wiggles". This is a major benefit of combining models in this manner as it overcomes the discreteness that occurs because of the number of components. The posterior distributions of the effective radii and flux for the different models, along with the combined distribution are compared in Figure~\ref{fig:comb_params}. We find that the distributions for all the models are similar, even those with lower evidence.

\begin{figure*}
    \centering
    \includegraphics[width = \textwidth]{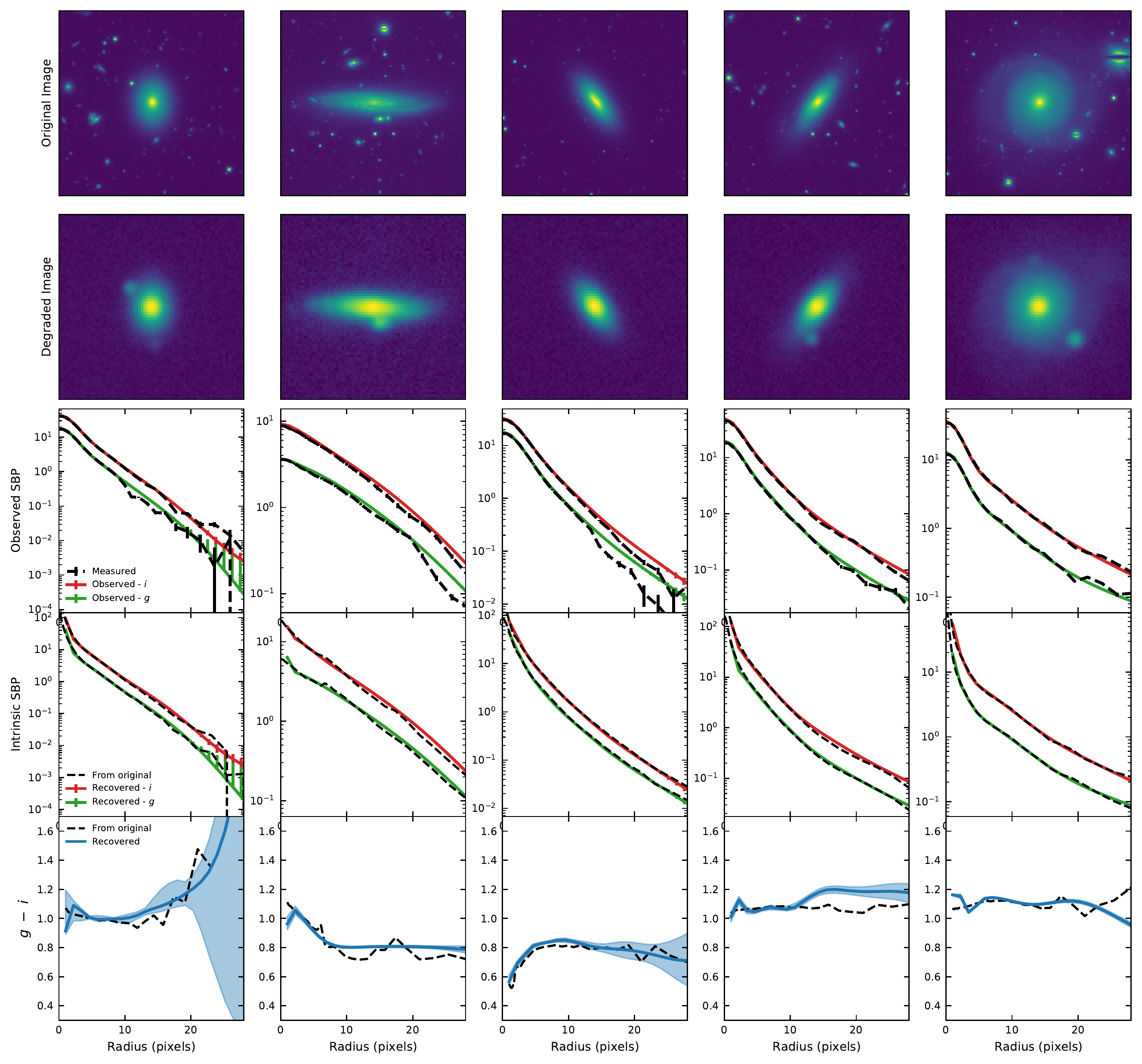}
    \caption{Five examples of local galaxies degraded and fit with \imc~ are shown. \textit{top:} The original $i$ HSC band images downloaded from the PDR2 archive are shown, each cutout is $\sim 1\arcmin \times 1\arcmin$. \textit{2nd row:} Degraded versions of the original images (see text). \textit{3rd row:} \imc~fits to the degraded images in the $i$ and $g$ bands are shown. The observed surface brightness profiles are from the combined distribution of five models with the errorbars showing the 16th-84th percentile range. These are compared to the surface brightness profile measured directly from the degraded images.\textit{4th row:} The \imc~recovered intrinsic surface brightness profile is shown. It is compared to the profile measured in the original image in elliptical apertures, following the geometry of the best fit \imc model. Even after the degradation \imc~ accurately recovers the original profile for all the galaxies.\textit{Bottom Row:} The intrinsic color profiles measured with \imc~ compared to those measured directly from the original images. In general we show \imc~can recover the surface brightness and color profiles for a large range of complex galaxy profiles.}
    \label{fig:degrad_all}
\end{figure*}

In Figure~\ref{fig:degrad_all} we show all five galaxies that have been degraded and fit with \imc. We show the recovered $g$ and $i$ band profiles along with the $g-i$ color. We measure the original surface brightness profiles in elliptical apertures following the best-fit geometry of the \imc fit. Each galaxy is fit with five models, with the same set up discussed and shown in Fig~\ref{fig:example_fit} and Fig.~\ref{fig:comb_sbp}, and combined using Bayesian model averaging. The errors bars on the recovered surface brightness profiles and the range shown in the color is the 5th-95th percentile.

\imc~is able to  recover the complex profiles of all five galaxies shown. We show a diversity of profiles with a disky galaxy in the second row and galaxies with multiple visible components in rows one and five. Along with the surface brightness profiles, our method recovers accurate color profiles. It is able to recover accurate profiles for those which have large slopes near the center, such as row three,  more gradual gradients like in row two and five, and flat color profiles, such as row 4. There are some parts of the profile where the recovered profile lies outside the \imc\ error ranges. This is due to the fact that a series of Gaussians can not fully model the galaxy. In \imc\ non-monotonic features, bumps or sharp breaks in the outer profile, where there are often less components delegated, often cannot be modelled accurately. However, all the large scale gradients and the bulk properties are recovered accurately. The flexibility of the MGE approach allows us to accurately recover the diversity of galaxy profiles shown here.

\section{Discussion and Summary}
\label{sec:summary}

In this paper we have described a Bayesian approach to fitting MGE models to images of galaxies and other astronomical objects. MGE represents galaxies by using a series of 2D Gaussians and is a flexible approach to studying galaxy morphology. Combined with Bayesian techniques this should help overcome some of the biases incurred by fitting traditional parameterized models to galaxies and derive accurate uncertainties on morphological parameters. Additionally, by using a MGE model of the PSF, convolution can be done analytically forgoing the traditional numerical approach.  We describe a technique for $\chi^2$ minimization along with a novel Bayesian inference method using pre-rendered images that is orders of magnitudes quicker then traditional rendering. We show a series of validation tests including injection-recovery tests. By injecting $z\sim1.5$ like galaxies in deep HST data we show that it can accurately recover morphological quantities. Additionally we show examples of our method using degraded images of local galaxies; this demonstrates that our method can recover profiles of realistic galaxies. We release a open-source python implementation of this method available via github\footnote{https://github.com/tbmiller-astro/imcascade}.

\imc is a Bayesian approach for fitting MGE models to galaxy images. It represents an alternative to the \texttt{mgefit} algorithm described in  \citetalias{Cappellari2002}. The major difference is that \texttt{mgefit} bins pixels in sectors before fitting whereas \imc~compares data to model pixel by pixel. The obvious advantage of the \texttt{mgefit} approach is computational efficiency. For an exponential profile placed into a 150 by 150 pixel cutout using nine Gaussian components, \texttt{mgefit} takes an average of 2.3 seconds to find a best fit model, and only marginally longer for a 300 by 300 pixel images, with an average run time of 2.6 seconds. With the same setup,\imc~ takes an average of 8.6 seconds and 30 seconds to find the model that minimizes $\chi^2$, respectively for the same size images\footnote{All tests were run on a desktop with a Intel Core i7-4770 CPU}. The run time scales nearly as the number of pixels as one would naively expect. The benefit of pixel-by-pixel comparison is that the full information in the image it utilized and one can adopt a Bayesian approach where the likelihood is calculated from the pixel fluxes and their associated weights. 

MGE models, implemented in \imc\, provide an alternative to traditional parameterization of galaxy profiles such as Sersic, or bulge+disk models. The flexibility of MGE means that there is no need specify a parameterization a-priori, like a.  While it may be possible to provide as good a fit to many galaxies, using multiple Sersic profiles for example, it often requires a lot of manual intervention as it is non-trivial to decide how many components to use. An additional benefit is is the ability to naturally fit the low surface-brightness outskirts of galaxies which are difficult to fit with paramaterized models. Due to their low S/N ratio, the $\chi^2$ and best-fit parameters are often dominated by the inner parts. In an MGE approach, the outskirts are fit by the components with the largest widths which are de-coupled from the inner components. Using a Bayesian method to estimate the posterior is also important in this regime as it provides a better estimate on the uncertainties of light in the low signal to noise regime. Finally the flexible MGE approach combined with Bayesian inference allows us the calculate reliable uncertainties on morphological quantities, which is often difficult with parameterized techniques as the systematic effects are unknown.

However, this leads to a philosophical departure from many previous studies which view galaxies through the lens of multiple discrete components~\citep[e.g.][]{Lackner2012,Tacchella2019}. While this is a useful approach when the galactic components can be reasonably separated and the formation scenarios are thought to be distinct, there are benefits to a more flexible, albeit less physically motivated approach~\citep{Steinmetz2002}. First off it is simpler and more direct to compare observation with simulations or semi-analytic models. Similar to observations it is non-trivial to separate galaxies into different components in simulations~\citep{Objera2018,Jagvaral2021} and the methods used are different then in observations. In the MGE framework it is simple to compare mass or light profiles or focus on specific parts of the galaxy that are of interest like the central regions or outskirts. However, many traditional parameterized measures of galaxies, such as the bulge mass, or Sersic index need to be replace with similar measures such as the mass within $1\ \rm kpc$ or the concentration ($5\log (r_{80}/r_{20})$), respectively, for example.~\citep{Cappellari2013}

There are additional features that we plan to add to \imc~which are useful for certain applications. One feature is the ability to fit two images simultaneously, for example high resolution HST and lower resolution ground based images. This feature, implemented in \texttt{mgefit} allows the total flux on large scale to be constrained by the lower resolution data while the internal structure is constrained by the high-resolution image. Additionally it is possible to implement a scheme where the axis ratio or position angle varies between the Gaussian components~\citepalias{Cappellari2002}. This allows more flexibility to encompass isophotal twists or changes in the axis ratio that occur in realistic galaxies. Finally, another feature currently employed in other methods is the ability to fit multiple objects in the same image simultaneously ~\citep{Peng2010,Erwin2015}. This is particularly useful for deblending objects that overlap on the sky.
\vspace{0.5mm}

\acknowledgments
We thank the anonymous referee for their comments which improved this manuscript. The authors would like to thank Imad Pasha and Josh Speagle for useful discussion on the name, design and implementation of \imc. TBM would like to thank Patricia Gruber and the Gruber foundation for their support of the work presented here. GAMA is a joint European-Australasian project based around a spectroscopic campaign using the Anglo-Australian Telescope. The GAMA input catalogue is based on data taken from the Sloan Digital Sky Survey and the UKIRT Infrared Deep Sky Survey. Complementary imaging of the GAMA regions is being obtained by a number of independent survey programmes including GALEX MIS, VST KiDS, VISTA VIKING, WISE, Herschel-ATLAS, GMRT and ASKAP providing UV to radio coverage. GAMA is funded by the STFC (UK), the ARC (Australia), the AAO, and the participating institutions. The GAMA website is http://www.gama-survey.org/.
\vspace{0.5mm}
\software{\texttt{numpy} \citep{numpy}, \texttt{scipy} \citep{scipy}, matplotlib \citep{matplotlib}, \texttt{astropy} \citep{Astropy2018}, \texttt{dynesty} \citep{dynesty}, photutils \citep{photutils}, Pandas \citet{pandas}, asdf \citep{asdf},\texttt{cornerpy} \citep{corner}, \texttt{SEP} \citep{SEP,SExtractor}, \texttt{mgefit}\citepalias{Cappellari2002}, \texttt{imcascade} \citep{imc_zenodo} }
\vspace{0.5mm}

\bibliography{imcascade,software}

\begin{thebibliography}{}
\expandafter\ifx\csname natexlab\endcsname\relax\def\natexlab#1{#1}\fi
\providecommand{\url}[1]{\href{#1}{#1}}
\providecommand{\dodoi}[1]{doi:~\href{http://doi.org/#1}{\nolinkurl{#1}}}
\providecommand{\doeprint}[1]{\href{http://ascl.net/#1}{\nolinkurl{http://ascl.net/#1}}}
\providecommand{\doarXiv}[1]{\href{https://arxiv.org/abs/#1}{\nolinkurl{https://arxiv.org/abs/#1}}}

\bibitem[{{Abraham} {et~al.}(1996){Abraham}, {Tanvir}, {Santiago}, {Ellis},
  {Glazebrook}, \& {van den Bergh}}]{Abraham1996}
{Abraham}, R.~G., {Tanvir}, N.~R., {Santiago}, B.~X., {et~al.} 1996, \mnras,
  279, L47, \dodoi{10.1093/mnras/279.3.L47}

\bibitem[{{Aihara} {et~al.}(2019){Aihara}, {AlSayyad}, {Ando}, {Armstrong},
  {Bosch}, {Egami}, {Furusawa}, {Furusawa}, {Goulding}, {Harikane}, {Hikage},
  {Ho}, {Hsieh}, {Huang}, {Ikeda}, {Imanishi}, {Ito}, {Iwata}, {Jaelani},
  {Kakuma}, {Kawana}, {Kikuta}, {Kobayashi}, {Koike}, {Komiyama}, {Li},
  {Liang}, {Lin}, {Luo}, {Lupton}, {Lust}, {MacArthur}, {Matsuoka}, {Mineo},
  {Miyatake}, {Miyazaki}, {More}, {Murata}, {Namiki}, {Nishizawa}, {Oguri},
  {Okabe}, {Okamoto}, {Okura}, {Ono}, {Onodera}, {Onoue}, {Osato}, {Ouchi},
  {Shibuya}, {Strauss}, {Sugiyama}, {Suto}, {Takada}, {Takagi}, {Takata},
  {Takita}, {Tanaka}, {Terai}, {Toba}, {Uchiyama}, {Utsumi}, {Wang}, {Wang}, \&
  {Yamada}}]{Aihara2019}
{Aihara}, H., {AlSayyad}, Y., {Ando}, M., {et~al.} 2019, \pasj, 71, 114,
  \dodoi{10.1093/pasj/psz103}

\bibitem[{{Astropy Collaboration} {et~al.}(2018){Astropy Collaboration},
  {Price-Whelan}, {Sip{\H{o}}cz}, {G{\"u}nther}, {Lim}, {Crawford}, {Conseil},
  {Shupe}, {Craig}, {Dencheva}, {Ginsburg}, {Vand erPlas}, {Bradley},
  {P{\'e}rez-Su{\'a}rez}, {de Val-Borro}, {Aldcroft}, {Cruz}, {Robitaille},
  {Tollerud}, {Ardelean}, {Babej}, {Bach}, {Bachetti}, {Bakanov}, {Bamford},
  {Barentsen}, {Barmby}, {Baumbach}, {Berry}, {Biscani}, {Boquien}, {Bostroem},
  {Bouma}, {Brammer}, {Bray}, {Breytenbach}, {Buddelmeijer}, {Burke},
  {Calderone}, {Cano Rodr{\'\i}guez}, {Cara}, {Cardoso}, {Cheedella}, {Copin},
  {Corrales}, {Crichton}, {D'Avella}, {Deil}, {Depagne}, {Dietrich}, {Donath},
  {Droettboom}, {Earl}, {Erben}, {Fabbro}, {Ferreira}, {Finethy}, {Fox},
  {Garrison}, {Gibbons}, {Goldstein}, {Gommers}, {Greco}, {Greenfield},
  {Groener}, {Grollier}, {Hagen}, {Hirst}, {Homeier}, {Horton}, {Hosseinzadeh},
  {Hu}, {Hunkeler}, {Ivezi{\'c}}, {Jain}, {Jenness}, {Kanarek}, {Kendrew},
  {Kern}, {Kerzendorf}, {Khvalko}, {King}, {Kirkby}, {Kulkarni}, {Kumar},
  {Lee}, {Lenz}, {Littlefair}, {Ma}, {Macleod}, {Mastropietro}, {McCully},
  {Montagnac}, {Morris}, {Mueller}, {Mumford}, {Muna}, {Murphy}, {Nelson},
  {Nguyen}, {Ninan}, {N{\"o}the}, {Ogaz}, {Oh}, {Parejko}, {Parley}, {Pascual},
  {Patil}, {Patil}, {Plunkett}, {Prochaska}, {Rastogi}, {Reddy Janga},
  {Sabater}, {Sakurikar}, {Seifert}, {Sherbert}, {Sherwood-Taylor}, {Shih},
  {Sick}, {Silbiger}, {Singanamalla}, {Singer}, {Sladen}, {Sooley},
  {Sornarajah}, {Streicher}, {Teuben}, {Thomas}, {Tremblay}, {Turner},
  {Terr{\'o}n}, {van Kerkwijk}, {de la Vega}, {Watkins}, {Weaver}, {Whitmore},
  {Woillez}, {Zabalza}, \& {Astropy Contributors}}]{Astropy2018}
{Astropy Collaboration}, {Price-Whelan}, A.~M., {Sip{\H{o}}cz}, B.~M., {et~al.}
  2018, \aj, 156, 123, \dodoi{10.3847/1538-3881/aabc4f}

\bibitem[{Barbary(2016)}]{SEP}
Barbary, K. 2016, Journal of Open Source Software, 1, 58,
  \dodoi{10.21105/joss.00058}

\bibitem[{{Bendinelli}(1991)}]{Bendinelli1991}
{Bendinelli}, O. 1991, \apj, 366, 599, \dodoi{10.1086/169595}

\bibitem[{{Bertin} \& {Arnouts}(1996)}]{SExtractor}
{Bertin}, E., \& {Arnouts}, S. 1996, \aaps, 117, 393,
  \dodoi{10.1051/aas:1996164}

\bibitem[{{Blanton} {et~al.}(2001){Blanton}, {Dalcanton}, {Eisenstein},
  {Loveday}, {Strauss}, {SubbaRao}, {Weinberg}, {Anderson}, {Annis}, {Bahcall},
  {Bernardi}, {Brinkmann}, {Brunner}, {Burles}, {Carey}, {Castander},
  {Connolly}, {Csabai}, {Doi}, {Finkbeiner}, {Friedman}, {Frieman}, {Fukugita},
  {Gunn}, {Hennessy}, {Hindsley}, {Hogg}, {Ichikawa}, {Ivezi{\'c}}, {Kent},
  {Knapp}, {Lamb}, {Leger}, {Long}, {Lupton}, {McKay}, {Meiksin}, {Merelli},
  {Munn}, {Narayanan}, {Newcomb}, {Nichol}, {Okamura}, {Owen}, {Pier}, {Pope},
  {Postman}, {Quinn}, {Rockosi}, {Schlegel}, {Schneider}, {Shimasaku},
  {Siegmund}, {Smee}, {Snir}, {Stoughton}, {Stubbs}, {Szalay}, {Szokoly},
  {Thakar}, {Tremonti}, {Tucker}, {Uomoto}, {Vanden Berk}, {Vogeley},
  {Waddell}, {Yanny}, {Yasuda}, \& {York}}]{Blanton2001}
{Blanton}, M.~R., {Dalcanton}, J., {Eisenstein}, D., {et~al.} 2001, \aj, 121,
  2358, \dodoi{10.1086/320405}

\bibitem[{Bradley {et~al.}(2020)Bradley, Sipőcz, Robitaille, Tollerud,
  Vinícius, Deil, Barbary, Wilson, Busko, Günther, Cara, Conseil, Bostroem,
  Droettboom, Bray, Bratholm, Lim, Barentsen, Craig, Pascual, Perren, Greco,
  Donath, de~Val-Borro, Kerzendorf, Bach, Weaver, D'Eugenio, Souchereau, \&
  Ferreira}]{photutils}
Bradley, L., Sipőcz, B., Robitaille, T., {et~al.} 2020, astropy/photutils:
  1.0.0, 1.0.0,  Zenodo, \dodoi{10.5281/zenodo.4044744}

\bibitem[{Branch {et~al.}(1999)Branch, Coleman, \& Li}]{trf}
Branch, M.~A., Coleman, T.~F., \& Li, Y. 1999, SIAM Journal on Scientific
  Computing, 21, \dodoi{10.1137/S1064827595289108}

\bibitem[{{Bundy} {et~al.}(2012){Bundy}, {Hogg}, {Higgs}, {Nichol}, {Yasuda},
  {Masters}, {Lang}, \& {Wake}}]{Bundy2012}
{Bundy}, K., {Hogg}, D.~W., {Higgs}, T.~D., {et~al.} 2012, \aj, 144, 188,
  \dodoi{10.1088/0004-6256/144/6/188}

\bibitem[{{Cappellari}(2002)}]{Cappellari2002}
{Cappellari}, M. 2002, \mnras, 333, 400,
  \dodoi{10.1046/j.1365-8711.2002.05412.x}

\bibitem[{{Cappellari}(2008)}]{Cappellari2008}
---. 2008, \mnras, 390, 71, \dodoi{10.1111/j.1365-2966.2008.13754.x}

\bibitem[{{Cappellari} {et~al.}(2002){Cappellari}, {Verolme}, {van der Marel},
  {Verdoes Kleijn}, {Illingworth}, {Franx}, {Carollo}, \& {de
  Zeeuw}}]{Cappellari2002a}
{Cappellari}, M., {Verolme}, E.~K., {van der Marel}, R.~P., {et~al.} 2002,
  \apj, 578, 787, \dodoi{10.1086/342653}

\bibitem[{{Cappellari} {et~al.}(2012){Cappellari}, {McDermid}, {Alatalo},
  {Blitz}, {Bois}, {Bournaud}, {Bureau}, {Crocker}, {Davies}, {Davis}, {de
  Zeeuw}, {Duc}, {Emsellem}, {Khochfar}, {Krajnovi{\'c}}, {Kuntschner},
  {Lablanche}, {Morganti}, {Naab}, {Oosterloo}, {Sarzi}, {Scott}, {Serra},
  {Weijmans}, \& {Young}}]{Cappellari2012}
{Cappellari}, M., {McDermid}, R.~M., {Alatalo}, K., {et~al.} 2012, \nat, 484,
  485, \dodoi{10.1038/nature10972}

\bibitem[{{Cappellari} {et~al.}(2013){Cappellari}, {Scott}, {Alatalo}, {Blitz},
  {Bois}, {Bournaud}, {Bureau}, {Crocker}, {Davies}, {Davis}, {de Zeeuw},
  {Duc}, {Emsellem}, {Khochfar}, {Krajnovi{\'c}}, {Kuntschner}, {McDermid},
  {Morganti}, {Naab}, {Oosterloo}, {Sarzi}, {Serra}, {Weijmans}, \&
  {Young}}]{Cappellari2013}
{Cappellari}, M., {Scott}, N., {Alatalo}, K., {et~al.} 2013, \mnras, 432, 1709,
  \dodoi{10.1093/mnras/stt562}

\bibitem[{Carollo {et~al.}(2013)Carollo, Bschorr, Renzini, Lilly, Capak,
  Cibinel, Ilbert, Onodera, Scoville, Cameron, Mobasher, Sanders, \&
  Taniguchi}]{Carollo2013}
Carollo, C.~M., Bschorr, T.~J., Renzini, A., {et~al.} 2013, Astrophysical
  Journal, 773

\bibitem[{{Conselice}(2003)}]{Conselice2003}
{Conselice}, C.~J. 2003, \apjs, 147, 1, \dodoi{10.1086/375001}

\bibitem[{{Conselice}(2014)}]{Conselice2014}
---. 2014, \araa, 52, 291, \dodoi{10.1146/annurev-astro-081913-040037}

\bibitem[{{de Souza} {et~al.}(2004){de Souza}, {Gadotti}, \& {dos
  Anjos}}]{deSouza2004}
{de Souza}, R.~E., {Gadotti}, D.~A., \& {dos Anjos}, S. 2004, \apjs, 153, 411,
  \dodoi{10.1086/421554}

\bibitem[{{D'Eugenio} {et~al.}(2021){D'Eugenio}, {Colless}, {Scott}, {van der
  Wel}, {Davies}, {van de Sande}, {Sweet}, {Oh}, {Groves}, {Sharp}, {Owers},
  {Bland-Hawthorn}, {Croom}, {Brough}, {Bryant}, {Goodwin}, {Lawrence},
  {Lorente}, \& {Richards}}]{Deugenio2021}
{D'Eugenio}, F., {Colless}, M., {Scott}, N., {et~al.} 2021, \mnras, 504, 5098,
  \dodoi{10.1093/mnras/stab1146}

\bibitem[{{Driver} {et~al.}(2011){Driver}, {Hill}, {Kelvin}, {Robotham},
  {Liske}, {Norberg}, {Baldry}, {Bamford}, {Hopkins}, {Loveday}, {Peacock},
  {Andrae}, {Bland-Hawthorn}, {Brough}, {Brown}, {Cameron}, {Ching}, {Colless},
  {Conselice}, {Croom}, {Cross}, {de Propris}, {Dye}, {Drinkwater}, {Ellis},
  {Graham}, {Grootes}, {Gunawardhana}, {Jones}, {van Kampen}, {Maraston},
  {Nichol}, {Parkinson}, {Phillipps}, {Pimbblet}, {Popescu}, {Prescott},
  {Roseboom}, {Sadler}, {Sansom}, {Sharp}, {Smith}, {Taylor}, {Thomas},
  {Tuffs}, {Wijesinghe}, {Dunne}, {Frenk}, {Jarvis}, {Madore}, {Meyer},
  {Seibert}, {Staveley-Smith}, {Sutherland}, \& {Warren}}]{Driver2011}
{Driver}, S.~P., {Hill}, D.~T., {Kelvin}, L.~S., {et~al.} 2011, \mnras, 413,
  971, \dodoi{10.1111/j.1365-2966.2010.18188.x}

\bibitem[{{Emsellem} {et~al.}(1994{\natexlab{a}}){Emsellem}, {Monnet}, \&
  {Bacon}}]{Emsellem1994a}
{Emsellem}, E., {Monnet}, G., \& {Bacon}, R. 1994{\natexlab{a}}, \aap, 285, 723

\bibitem[{{Emsellem} {et~al.}(1994{\natexlab{b}}){Emsellem}, {Monnet}, {Bacon},
  \& {Nieto}}]{Emsellem1994b}
{Emsellem}, E., {Monnet}, G., {Bacon}, R., \& {Nieto}, J.~L.
  1994{\natexlab{b}}, \aap, 285, 739

\bibitem[{{Erwin}(2015)}]{Erwin2015}
{Erwin}, P. 2015, \apj, 799, 226, \dodoi{10.1088/0004-637X/799/2/226}

\bibitem[{Foreman-Mackey(2016)}]{corner}
Foreman-Mackey, D. 2016, The Journal of Open Source Software, 1, 24,
  \dodoi{10.21105/joss.00024}

\bibitem[{Fragoso {et~al.}(2018)Fragoso, Bertoli, \& Louzada}]{Fragoso2018}
Fragoso, T.~M., Bertoli, W., \& Louzada, F. 2018, International Statistical
  Review, 86, 1, \dodoi{https://doi.org/10.1111/insr.12243}

\bibitem[{Fruhwirth-Schnatter {et~al.}(2019)Fruhwirth-Schnatter, Celeux, \&
  Robert}]{MMbook}
Fruhwirth-Schnatter, S., Celeux, G., \& Robert, C. 2019, Handbook of Mixture
  Analysis, Chapman \& Hall/CRC Handbooks of Modern Statistical Methods (CRC
  Press).
\newblock \url{https://books.google.com/books?id=Hu-yDwAAQBAJ}

\bibitem[{Gelman {et~al.}(2013)Gelman, Carlin, Stern, \& Rubin}]{Gelman2013}
Gelman, A., Carlin, J.~B., Stern, H.~S., \& Rubin, D.~B. 2013, Bayesian Data
  Analysis, 3rd edn. (Chapman and Hall/CRC)

\bibitem[{Greenfield {et~al.}(2015)Greenfield, Droettboom, \& Bray}]{asdf}
Greenfield, P., Droettboom, M., \& Bray, E. 2015, Astronomy and Computing, 12,
  240, \dodoi{https://doi.org/10.1016/j.ascom.2015.06.004}

\bibitem[{{Grogin} {et~al.}(2011){Grogin}, {Kocevski}, {Faber}, {Ferguson},
  {Koekemoer}, {Riess}, {Acquaviva}, {Alexander}, {Almaini}, {Ashby}, {Barden},
  {Bell}, {Bournaud}, {Brown}, {Caputi}, {Casertano}, {Cassata}, {Castellano},
  {Challis}, {Chary}, {Cheung}, {Cirasuolo}, {Conselice}, {Roshan Cooray},
  {Croton}, {Daddi}, {Dahlen}, {Dav{\'e}}, {de Mello}, {Dekel}, {Dickinson},
  {Dolch}, {Donley}, {Dunlop}, {Dutton}, {Elbaz}, {Fazio}, {Filippenko},
  {Finkelstein}, {Fontana}, {Gardner}, {Garnavich}, {Gawiser}, {Giavalisco},
  {Grazian}, {Guo}, {Hathi}, {H{\"a}ussler}, {Hopkins}, {Huang}, {Huang},
  {Jha}, {Kartaltepe}, {Kirshner}, {Koo}, {Lai}, {Lee}, {Li}, {Lotz}, {Lucas},
  {Madau}, {McCarthy}, {McGrath}, {McIntosh}, {McLure}, {Mobasher},
  {Moustakas}, {Mozena}, {Nandra}, {Newman}, {Niemi}, {Noeske}, {Papovich},
  {Pentericci}, {Pope}, {Primack}, {Rajan}, {Ravindranath}, {Reddy}, {Renzini},
  {Rix}, {Robaina}, {Rodney}, {Rosario}, {Rosati}, {Salimbeni}, {Scarlata},
  {Siana}, {Simard}, {Smidt}, {Somerville}, {Spinrad}, {Straughn}, {Strolger},
  {Telford}, {Teplitz}, {Trump}, {van der Wel}, {Villforth}, {Wechsler},
  {Weiner}, {Wiklind}, {Wild}, {Wilson}, {Wuyts}, {Yan}, \& {Yun}}]{Grogin2011}
{Grogin}, N.~A., {Kocevski}, D.~D., {Faber}, S.~M., {et~al.} 2011, \apjs, 197,
  35, \dodoi{10.1088/0067-0049/197/2/35}

\bibitem[{{Higson} {et~al.}(2019){Higson}, {Handley}, {Hobson}, \&
  {Lasenby}}]{Higson2019}
{Higson}, E., {Handley}, W., {Hobson}, M., \& {Lasenby}, A. 2019, Statistics
  and Computing, 29, 891, \dodoi{10.1007/s11222-018-9844-0}

\bibitem[{{Hogg} \& {Lang}(2013)}]{Hogg2013}
{Hogg}, D.~W., \& {Lang}, D. 2013, \pasp, 125, 719, \dodoi{10.1086/671228}

\bibitem[{{Holmberg}(1958)}]{Holmberg1958}
{Holmberg}, E. 1958, Meddelanden fran Lunds Astronomiska Observatorium Serie
  II, 136, 1

\bibitem[{Hunter(2007)}]{matplotlib}
Hunter, J.~D. 2007, Computing in Science \& Engineering, 9, 90,
  \dodoi{10.1109/MCSE.2007.55}

\bibitem[{{Jagvaral} {et~al.}(2021){Jagvaral}, {Campbell}, {Mandelbaum}, \&
  {Rau}}]{Jagvaral2021}
{Jagvaral}, Y., {Campbell}, D., {Mandelbaum}, R., \& {Rau}, M.~M. 2021, arXiv
  e-prints, arXiv:2105.02237.
\newblock \doarXiv{2105.02237}

\bibitem[{{Jedrzejewski}(1987)}]{ellipse}
{Jedrzejewski}, R.~I. 1987, \mnras, 226, 747, \dodoi{10.1093/mnras/226.4.747}

\bibitem[{Kass \& Raftery(1995)}]{Kass1995}
Kass, R.~E., \& Raftery, A.~E. 1995, Journal of the American Statistical
  Association, 90, 773, \dodoi{10.1080/01621459.1995.10476572}

\bibitem[{{Koekemoer} {et~al.}(2011){Koekemoer}, {Faber}, {Ferguson}, {Grogin},
  {Kocevski}, {Koo}, {Lai}, {Lotz}, {Lucas}, {McGrath}, {Ogaz}, {Rajan},
  {Riess}, {Rodney}, {Strolger}, {Casertano}, {Castellano}, {Dahlen},
  {Dickinson}, {Dolch}, {Fontana}, {Giavalisco}, {Grazian}, {Guo}, {Hathi},
  {Huang}, {van der Wel}, {Yan}, {Acquaviva}, {Alexander}, {Almaini}, {Ashby},
  {Barden}, {Bell}, {Bournaud}, {Brown}, {Caputi}, {Cassata}, {Challis},
  {Chary}, {Cheung}, {Cirasuolo}, {Conselice}, {Roshan Cooray}, {Croton},
  {Daddi}, {Dav{\'e}}, {de Mello}, {de Ravel}, {Dekel}, {Donley}, {Dunlop},
  {Dutton}, {Elbaz}, {Fazio}, {Filippenko}, {Finkelstein}, {Frazer}, {Gardner},
  {Garnavich}, {Gawiser}, {Gruetzbauch}, {Hartley}, {H{\"a}ussler},
  {Herrington}, {Hopkins}, {Huang}, {Jha}, {Johnson}, {Kartaltepe},
  {Khostovan}, {Kirshner}, {Lani}, {Lee}, {Li}, {Madau}, {McCarthy},
  {McIntosh}, {McLure}, {McPartland}, {Mobasher}, {Moreira}, {Mortlock},
  {Moustakas}, {Mozena}, {Nandra}, {Newman}, {Nielsen}, {Niemi}, {Noeske},
  {Papovich}, {Pentericci}, {Pope}, {Primack}, {Ravindranath}, {Reddy},
  {Renzini}, {Rix}, {Robaina}, {Rosario}, {Rosati}, {Salimbeni}, {Scarlata},
  {Siana}, {Simard}, {Smidt}, {Snyder}, {Somerville}, {Spinrad}, {Straughn},
  {Telford}, {Teplitz}, {Trump}, {Vargas}, {Villforth}, {Wagner}, {Wandro},
  {Wechsler}, {Weiner}, {Wiklind}, {Wild}, {Wilson}, {Wuyts}, \&
  {Yun}}]{Koekemoer2011}
{Koekemoer}, A.~M., {Faber}, S.~M., {Ferguson}, H.~C., {et~al.} 2011, \apjs,
  197, 36, \dodoi{10.1088/0067-0049/197/2/36}

\bibitem[{{Kron}(1980)}]{Kron1980}
{Kron}, R.~G. 1980, \apjs, 43, 305, \dodoi{10.1086/190669}

\bibitem[{{Lackner} \& {Gunn}(2012)}]{Lackner2012}
{Lackner}, C.~N., \& {Gunn}, J.~E. 2012, \mnras, 421, 2277,
  \dodoi{10.1111/j.1365-2966.2012.20450.x}

\bibitem[{{Lang}(2020)}]{Lang2020}
{Lang}, D. 2020, arXiv e-prints, arXiv:2012.15797.
\newblock \doarXiv{2012.15797}

\bibitem[{{Lang} {et~al.}(2014){Lang}, {Wuyts}, {Somerville}, {F{\"o}rster
  Schreiber}, {Genzel}, {Bell}, {Brammer}, {Dekel}, {Faber}, {Ferguson},
  {Grogin}, {Kocevski}, {Koekemoer}, {Lutz}, {McGrath}, {Momcheva}, {Nelson},
  {Primack}, {Rosario}, {Skelton}, {Tacconi}, {van Dokkum}, \&
  {Whitaker}}]{Lang2014}
{Lang}, P., {Wuyts}, S., {Somerville}, R.~S., {et~al.} 2014, \apj, 788, 11,
  \dodoi{10.1088/0004-637X/788/1/11}

\bibitem[{Lange {et~al.}(2015)Lange, Driver, Robotham, Kelvin, Graham,
  Alpaslan, Andrews, Baldry, Bamford, Bland-Hawthorn, Brough, Cluver,
  Conselice, Davies, Haeussler, Konstantopoulos, Loveday, Moffett, Norberg,
  Phillipps, Taylor, L{\'{o}}pez-S{\'{a}}nchez, \& Wilkins}]{Lange2015}
Lange, R., Driver, S.~P., Robotham, A.~S., {et~al.} 2015, Monthly Notices of
  the Royal Astronomical Society, 447, 2603

\bibitem[{{Li} {et~al.}(2018){Li}, {Mao}, {Cappellari}, {Ge}, {Long}, {Li},
  {Mo}, {Li}, {Zheng}, {Bundy}, {Thomas}, {Brownstein}, {Roman Lopes}, {Law},
  \& {Drory}}]{Hongyu2018}
{Li}, H., {Mao}, S., {Cappellari}, M., {et~al.} 2018, \mnras, 476, 1765,
  \dodoi{10.1093/mnras/sty334}

\bibitem[{{Mendel} {et~al.}(2020{\natexlab{a}}){Mendel}, {Beifiori}, {Saglia},
  {Bender}, {Brammer}, {Chan}, {F{\"o}rster Schreiber}, {Fossati}, {Galametz},
  {Momcheva}, {Nelson}, {Wilman}, \& {Wuyts}}]{Mendel2020}
{Mendel}, J.~T., {Beifiori}, A., {Saglia}, R.~P., {et~al.} 2020{\natexlab{a}},
  \apj, 899, 87, \dodoi{10.3847/1538-4357/ab9ffc}

\bibitem[{{Mendel} {et~al.}(2020{\natexlab{b}}){Mendel}, {Beifiori}, {Saglia},
  {Bender}, {Brammer}, {Chan}, {F{\"o}rster Schreiber}, {Fossati}, {Galametz},
  {Momcheva}, {Nelson}, {Wilman}, \& {Wuyts}}]{Trevor2020}
---. 2020{\natexlab{b}}, \apj, 899, 87, \dodoi{10.3847/1538-4357/ab9ffc}

\bibitem[{Miller \& van Dokkum(2021)}]{imc_zenodo}
Miller, T.~B., \& van Dokkum, P. 2021, imcascade, 1.0,  Zenodo,
  \dodoi{10.5281/zenodo.5516734}

\bibitem[{{Monnet} {et~al.}(1992){Monnet}, {Bacon}, \& {Emsellem}}]{Monnet1992}
{Monnet}, G., {Bacon}, R., \& {Emsellem}, E. 1992, \aap, 253, 366

\bibitem[{{Mowla} {et~al.}(2019){Mowla}, {van Dokkum}, {Brammer}, {Momcheva},
  {van der Wel}, {Whitaker}, {Nelson}, {Bezanson}, {Muzzin}, {Franx},
  {MacKenty}, {Leja}, {Kriek}, \& {Marchesini}}]{Mowla2019}
{Mowla}, L.~A., {van Dokkum}, P., {Brammer}, G.~B., {et~al.} 2019, \apj, 880,
  57, \dodoi{10.3847/1538-4357/ab290a}

\bibitem[{{Obreja} {et~al.}(2018){Obreja}, {Macci{\`o}}, {Moster}, {Dutton},
  {Buck}, {Stinson}, \& {Wang}}]{Objera2018}
{Obreja}, A., {Macci{\`o}}, A.~V., {Moster}, B., {et~al.} 2018, \mnras, 477,
  4915, \dodoi{10.1093/mnras/sty1022}

\bibitem[{Ono {et~al.}(2013)Ono, Ouchi, Curtis-Lake, Schenker, Ellis, McLure,
  Dunlop, Robertson, Koekemoer, Bowler, Rogers, Schneider, Charlot, Stark,
  Shimasaku, Furlanetto, \& Cirasuolo}]{Ono2013}
Ono, Y., Ouchi, M., Curtis-Lake, E., {et~al.} 2013, The Astrophysical Journal,
  777, 155

\bibitem[{{Peng} {et~al.}(2010){Peng}, {Ho}, {Impey}, \& {Rix}}]{Peng2010}
{Peng}, C.~Y., {Ho}, L.~C., {Impey}, C.~D., \& {Rix}, H.-W. 2010, \aj, 139,
  2097, \dodoi{10.1088/0004-6256/139/6/2097}

\bibitem[{{Petrosian}(1976)}]{Petrosian1976}
{Petrosian}, V. 1976, \apjl, 210, L53, \dodoi{10.1086/182301}

\bibitem[{{Redman}(1936)}]{Redman1936}
{Redman}, R.~O. 1936, \mnras, 96, 588, \dodoi{10.1093/mnras/96.6.588}

\bibitem[{Roberts(1965)}]{Roberts1965}
Roberts, H.~V. 1965, Journal of the American Statistical Association, 60, 50,
  \dodoi{10.1080/01621459.1965.10480774}

\bibitem[{{Robotham} {et~al.}(2017){Robotham}, {Taranu}, {Tobar}, {Moffett}, \&
  {Driver}}]{Robotham2017}
{Robotham}, A.~S.~G., {Taranu}, D.~S., {Tobar}, R., {Moffett}, A., \& {Driver},
  S.~P. 2017, \mnras, 466, 1513, \dodoi{10.1093/mnras/stw3039}

\bibitem[{{Sersic}(1968)}]{Sersic1968}
{Sersic}, J.~L. 1968, {Atlas de Galaxias Australes}

\bibitem[{{Shajib}(2019)}]{Shajib2019}
{Shajib}, A.~J. 2019, \mnras, 488, 1387, \dodoi{10.1093/mnras/stz1796}

\bibitem[{{Sheldon}(2014)}]{Sheldon2014}
{Sheldon}, E.~S. 2014, \mnras, 444, L25, \dodoi{10.1093/mnrasl/slu104}

\bibitem[{Shen {et~al.}(2003)Shen, Mo, White, Blanton, Kauffmann, Voges,
  Brinkmann, \& Csabai}]{Shen2003}
Shen, S., Mo, H.~J., White, S. D.~M., {et~al.} 2003, Monthly Notice of the
  Royal Astronomical Society, 343, 978

\bibitem[{{Shetty} \& {Cappellari}(2015)}]{Shetty2015}
{Shetty}, S., \& {Cappellari}, M. 2015, \mnras, 454, 1332,
  \dodoi{10.1093/mnras/stv1948}

\bibitem[{{Shibuya} {et~al.}(2021){Shibuya}, {Miura}, {Iwadate}, {Fujimoto},
  {Harikane}, {Toba}, {Umayahara}, \& {Ito}}]{Shibuya2021}
{Shibuya}, T., {Miura}, N., {Iwadate}, K., {et~al.} 2021, arXiv e-prints,
  arXiv:2106.03728.
\newblock \doarXiv{2106.03728}

\bibitem[{{Simard}(1998)}]{Simard1998}
{Simard}, L. 1998, in Astronomical Society of the Pacific Conference Series,
  Vol. 145, Astronomical Data Analysis Software and Systems VII, ed.
  R.~{Albrecht}, R.~N. {Hook}, \& H.~A. {Bushouse}, 108

\bibitem[{{Skelton} {et~al.}(2014){Skelton}, {Whitaker}, {Momcheva}, {Brammer},
  {van Dokkum}, {Labb{\'e}}, {Franx}, {van der Wel}, {Bezanson}, {Da Cunha},
  {Fumagalli}, {F{\"o}rster Schreiber}, {Kriek}, {Leja}, {Lundgren}, {Magee},
  {Marchesini}, {Maseda}, {Nelson}, {Oesch}, {Pacifici}, {Patel}, {Price},
  {Rix}, {Tal}, {Wake}, \& {Wuyts}}]{Skelton2014}
{Skelton}, R.~E., {Whitaker}, K.~E., {Momcheva}, I.~G., {et~al.} 2014, \apjs,
  214, 24, \dodoi{10.1088/0067-0049/214/2/24}

\bibitem[{Skilling(2006)}]{Skilling2006}
Skilling, J. 2006, Bayesian Analysis, 1, 833 , \dodoi{10.1214/06-BA127}

\bibitem[{{Speagle}(2020)}]{dynesty}
{Speagle}, J.~S. 2020, \mnras, 493, 3132, \dodoi{10.1093/mnras/staa278}

\bibitem[{{Steinmetz} \& {Navarro}(2002)}]{Steinmetz2002}
{Steinmetz}, M., \& {Navarro}, J.~F. 2002, \na, 7, 155,
  \dodoi{10.1016/S1384-1076(02)00102-1}

\bibitem[{{Stone} {et~al.}(2021){Stone}, {Arora}, {Courteau}, \&
  {Cuillandre}}]{Stone2021}
{Stone}, C., {Arora}, N., {Courteau}, S., \& {Cuillandre}, J.-C. 2021, arXiv
  e-prints, arXiv:2106.13809.
\newblock \doarXiv{2106.13809}

\bibitem[{{Suess} {et~al.}(2019){Suess}, {Kriek}, {Price}, \&
  {Barro}}]{Suess2019}
{Suess}, K.~A., {Kriek}, M., {Price}, S.~H., \& {Barro}, G. 2019, \apj, 877,
  103, \dodoi{10.3847/1538-4357/ab1bda}

\bibitem[{{Suess} {et~al.}(2021){Suess}, {Kriek}, {Price}, \&
  {Barro}}]{Suess2021}
---. 2021, arXiv e-prints, arXiv:2101.05820.
\newblock \doarXiv{2101.05820}

\bibitem[{{Szomoru} {et~al.}(2012){Szomoru}, {Franx}, \& {van
  Dokkum}}]{Szomoru2012}
{Szomoru}, D., {Franx}, M., \& {van Dokkum}, P.~G. 2012, \apj, 749, 121,
  \dodoi{10.1088/0004-637X/749/2/121}

\bibitem[{{Szomoru} {et~al.}(2013){Szomoru}, {Franx}, {van Dokkum}, {Trenti},
  {Illingworth}, {Labb{\'e}}, \& {Oesch}}]{Szomoru2013}
{Szomoru}, D., {Franx}, M., {van Dokkum}, P.~G., {et~al.} 2013, \apj, 763, 73,
  \dodoi{10.1088/0004-637X/763/2/73}

\bibitem[{{Szomoru} {et~al.}(2010){Szomoru}, {Franx}, {van Dokkum}, {Trenti},
  {Illingworth}, {Labb{\'e}}, {Bouwens}, {Oesch}, \& {Carollo}}]{Szomoru2010}
---. 2010, \apjl, 714, L244, \dodoi{10.1088/2041-8205/714/2/L244}

\bibitem[{{Tacchella} {et~al.}(2019){Tacchella}, {Diemer}, {Hernquist},
  {Genel}, {Marinacci}, {Nelson}, {Pillepich}, {Rodriguez-Gomez}, {Sales},
  {Springel}, \& {Vogelsberger}}]{Tacchella2019}
{Tacchella}, S., {Diemer}, B., {Hernquist}, L., {et~al.} 2019, \mnras, 487,
  5416, \dodoi{10.1093/mnras/stz1657}

\bibitem[{{Taylor} {et~al.}(2011){Taylor}, {Hopkins}, {Baldry}, {Brown},
  {Driver}, {Kelvin}, {Hill}, {Robotham}, {Bland -Hawthorn}, {Jones}, {Sharp},
  {Thomas}, {Liske}, {Loveday}, {Norberg}, {Peacock}, {Bamford}, {Brough},
  {Colless}, {Cameron}, {Conselice}, {Croom}, {Frenk}, {Gunawardhana},
  {Kuijken}, {Nichol}, {Parkinson}, {Phillipps}, {Pimbblet}, {Popescu},
  {Prescott}, {Sutherland}, {Tuffs}, {van Kampen}, \&
  {Wijesinghe}}]{Taylor2011}
{Taylor}, E.~N., {Hopkins}, A.~M., {Baldry}, I.~K., {et~al.} 2011, \mnras, 418,
  1587, \dodoi{10.1111/j.1365-2966.2011.19536.x}

\bibitem[{{Trujillo} {et~al.}(2020){Trujillo}, {Chamba}, \&
  {Knapen}}]{Trujillo2020}
{Trujillo}, I., {Chamba}, N., \& {Knapen}, J.~H. 2020, \mnras, 493, 87,
  \dodoi{10.1093/mnras/staa236}

\bibitem[{Trujillo {et~al.}(2006)Trujillo, Forster~Schreiber, Rudnick, Barden,
  Franx, Rix, Caldwell, McIntosh, Toft, Haussler, Zirm, van Dokkum, Labbe,
  Moorwood, Rottgering, van~der Wel, van~der Werf, \& van
  Starkenburg}]{Trujillo2006}
Trujillo, I., Forster~Schreiber, N.~M., Rudnick, G., {et~al.} 2006, The
  Astrophysical Journal, 650, 18

\bibitem[{Van Der~Walt {et~al.}(2011)Van Der~Walt, Colbert, \&
  Varoquaux}]{numpy}
Van Der~Walt, S., Colbert, S.~C., \& Varoquaux, G. 2011, Computing in Science
  \& Engineering, 13, 22

\bibitem[{{van der Wel} {et~al.}(2012){van der Wel}, {Bell}, {H{\"a}ussler},
  {McGrath}, {Chang}, {Guo}, {McIntosh}, {Rix}, {Barden}, {Cheung}, {Faber},
  {Ferguson}, {Galametz}, {Grogin}, {Hartley}, {Kartaltepe}, {Kocevski},
  {Koekemoer}, {Lotz}, {Mozena}, {Peth}, \& {Peng}}]{vanderWel2012}
{van der Wel}, A., {Bell}, E.~F., {H{\"a}ussler}, B., {et~al.} 2012, \apjs,
  203, 24, \dodoi{10.1088/0067-0049/203/2/24}

\bibitem[{{van der Wel} {et~al.}(2014){van der Wel}, {Franx}, {van Dokkum},
  {Skelton}, {Momcheva}, {Whitaker}, {Brammer}, {Bell}, {Rix}, {Wuyts},
  {Ferguson}, {Holden}, {Barro}, {Koekemoer}, {Chang}, {McGrath},
  {H{\"a}ussler}, {Dekel}, {Behroozi}, {Fumagalli}, {Leja}, {Lundgren},
  {Maseda}, {Nelson}, {Wake}, {Patel}, {Labb{\'e}}, {Faber}, {Grogin}, \&
  {Kocevski}}]{vanderWel2014}
{van der Wel}, A., {Franx}, M., {van Dokkum}, P.~G., {et~al.} 2014, \apj, 788,
  28, \dodoi{10.1088/0004-637X/788/1/28}

\bibitem[{{van Dokkum} {et~al.}(2010){van Dokkum}, {Whitaker}, {Brammer},
  {Franx}, {Kriek}, {Labb{\'e}}, {Marchesini}, {Quadri}, {Bezanson},
  {Illingworth}, {Muzzin}, {Rudnick}, {Tal}, \& {Wake}}]{vanDokkum2010}
{van Dokkum}, P.~G., {Whitaker}, K.~E., {Brammer}, G., {et~al.} 2010, \apj,
  709, 1018, \dodoi{10.1088/0004-637X/709/2/1018}

\bibitem[{van Dokkum {et~al.}(2015)van Dokkum, Nelson, Franx, Oesch, Momcheva,
  Brammer, Schreiber, Skelton, Whitaker, Wel, Bezanson, Fumagalli, Illingworth,
  Kriek, Leja, \& Wuyts}]{VanDokkum2015}
van Dokkum, P.~G., Nelson, E.~J., Franx, M., {et~al.} 2015, Astrophysical
  Journal, 813

\bibitem[{{Virtanen} {et~al.}(2020){Virtanen}, {Gommers}, {Oliphant},
  {Haberland}, {Reddy}, {Cournapeau}, {Burovski}, {Peterson}, {Weckesser},
  {Bright}, {van der Walt}, {Brett}, {Wilson}, {Jarrod Millman}, {Mayorov},
  {Nelson}, {Jones}, {Kern}, {Larson}, {Carey}, {Polat}, {Feng}, {Moore}, {Vand
  erPlas}, {Laxalde}, {Perktold}, {Cimrman}, {Henriksen}, {Quintero}, {Harris},
  {Archibald}, {Ribeiro}, {Pedregosa}, {van Mulbregt}, \&
  {Contributors}}]{scipy}
{Virtanen}, P., {Gommers}, R., {Oliphant}, T.~E., {et~al.} 2020, Nature
  Methods, 17, 261, \dodoi{https://doi.org/10.1038/s41592-019-0686-2}

\bibitem[{{W}es {M}c{K}inney(2010)}]{pandas}
{W}es {M}c{K}inney. 2010, in {P}roceedings of the 9th {P}ython in {S}cience
  {C}onference, ed. {S}t\'efan van~der {W}alt \& {J}arrod {M}illman, 56 -- 61,
  \dodoi{10.25080/Majora-92bf1922-00a}

\bibitem[{{Whitaker} {et~al.}(2015){Whitaker}, {Franx}, {Bezanson}, {Brammer},
  {van Dokkum}, {Kriek}, {Labb{\'e}}, {Leja}, {Momcheva}, {Nelson}, {Rigby},
  {Rix}, {Skelton}, {van der Wel}, \& {Wuyts}}]{Whitaker2015}
{Whitaker}, K.~E., {Franx}, M., {Bezanson}, R., {et~al.} 2015, \apjl, 811, L12,
  \dodoi{10.1088/2041-8205/811/1/L12}

\bibitem[{Williams {et~al.}(2010)Williams, Quadri, Franx, Van~Dokkum, Toft,
  Kriek, \& Labb{\'{e}}}]{Williams2010}
Williams, R.~J., Quadri, R.~F., Franx, M., {et~al.} 2010, Astrophysical
  Journal, 713, 738

\bibitem[{{Wuyts} {et~al.}(2013){Wuyts}, {F{\"o}rster Schreiber}, {Nelson},
  {van Dokkum}, {Brammer}, {Chang}, {Faber}, {Ferguson}, {Franx}, {Fumagalli},
  {Genzel}, {Grogin}, {Kocevski}, {Koekemoer}, {Lundgren}, {Lutz}, {McGrath},
  {Momcheva}, {Rosario}, {Skelton}, {Tacconi}, {van der Wel}, \&
  {Whitaker}}]{Wuyts2013}
{Wuyts}, S., {F{\"o}rster Schreiber}, N.~M., {Nelson}, E.~J., {et~al.} 2013,
  \apj, 779, 135, \dodoi{10.1088/0004-637X/779/2/135}

\bibitem[{{Zibetti} {et~al.}(2009){Zibetti}, {Charlot}, \& {Rix}}]{Zibetti2009}
{Zibetti}, S., {Charlot}, S., \& {Rix}, H.-W. 2009, \mnras, 400, 1181,
  \dodoi{10.1111/j.1365-2966.2009.15528.x}

\end{thebibliography}
\bibliographystyle{aasjournal}

\appendix
\section{Non-circular PSF}
\label{sec:NC_psf}
In this section we provide a slight modification to the derivation of the convolution of two non-circular Gaussian in \citet{Monnet1992}. In particular their derivation solves the deconvolution problem i.e. the solution is expressed in terms of the angle between the convolved Gaussian, which we denote $\psi$ and the unconvolved Gaussian, which we denote $\phi$. For our purposes of generating models, we are interested in the convolution, i.e. solving for the convovled position angle, $\psi$ ,in terms of the intrinsic position angle, $\phi$, and the position angle of the PSF, $\theta$. We modify the original derivation slightly to solve for $\psi$ directly below.

To begin we re-write Eqn. 75 in Appendix B of \citet{Monnet1992}.
\begin{equation}
    \label{eqn:abc_to_solve}
    \begin{aligned}
    0 &= (1-q_i^2)\sigma_i^2 \sin 2\phi^* + (1-q_j^2)\sigma_j^2 \sin 2\theta^*\ (a) \\
    \sigma_O^2 &= \sigma_i^2 ( \cos^2 \phi^* + q_i^2\,\sin^2 \phi^* ) +  \sigma_j^2 ( \cos^2 \theta^* + q_j^2\,\sin^2 \theta^* )\ (b)\\
    q_O^2\, \sigma_O^2 &= \sigma_i^2 ( \sin^2 \phi^* + q_i^2\,\cos^2 \phi^* ) +  \sigma_j^2 ( \sin^2 \theta^* + q_j^2\,\cos^2 \theta^* )\ (c)
    \end{aligned}
\end{equation}

Here $\phi^* = \phi - \psi$, where $\phi$ is the intrinsic angle of Gaussian to be convovled in the frame rotated by $\psi$. Similarly, $\theta^* = \theta - \psi$ where $\theta$ is the intrinsic angle of PSF to be convovled in the frame rotated by $\psi$. $\phi^*$ and $\theta^*$ are denoted as $\alpha$ and $\beta$ respectively in \citet{Monnet1992}.  We are interested in expressing the solution in terms of $\psi$. Using the relation $(a)$ above we, and expand $\phi^*$ and $\theta^*$ to solve for $\psi$ as,

\begin{equation}
    \label{eqn:psi}
    \tan 2\psi = \frac{(1-q_i^2)\sigma_i^2\, \sin 2\phi + (1-q_j^2)\sigma_j^2\, \sin 2\theta} { (1-q_i^2)\sigma_i^2\, \cos 2\phi + (1-q_j^2)\sigma_j^2\, \cos 2\theta}
\end{equation}

Once we have calculated $\psi$, we can then calculate $\sigma_O$, the width of the convolved Gaussian using relation $(b)$ in Eqn~\ref{eqn:abc_to_solve} and then finally $q_O$, the axis ratio of the convolved Gaussian, using relation $(c)$. These three parameters, along with the product of the weights give and analytic description for the convolution of two non-circular Gaussians. In practice, we only recommend using a non-circular PSF when needed, as it increases the computation time to render each model. In this scenario, every component of $O$ has a different position angle and therefore must be rotated separately before being combined, whereas in the case of a circular PSF, all components have the same position angle and can therefore be combined before that one image is rotated.

\end{document}